\documentclass[aps,pra,preprint,superscriptaddress]{revtex4-2}

\usepackage{graphicx,amsmath,amsfonts,amssymb, url}
\usepackage[utf8x]{inputenc}
\usepackage[T1]{fontenc}
\usepackage[breaklinks = true]{hyperref}
\usepackage[dvipsnames]{xcolor}

\hypersetup{
    colorlinks=true,
    linkcolor=blue,
    filecolor=magenta,      
    urlcolor=blue,
    citecolor=red
}

\graphicspath{ {./} }

\makeatletter

\newcommand\norm[1]{\left\lVert#1\right\rVert}
\newcommand{\twodots}{\mathinner {\ldotp \ldotp}}
\newcommand\new[1]{#1}

\begin{document}

\title{Learning and avoiding disorder in multimode fibers}

\author{Maxime W. Matth{\`e}s}
\affiliation{Institut Langevin, ESPCI Paris, PSL University, CNRS, France}
\author{Yaron Bromberg}
\affiliation{Racah Institute of Physics, The Hebrew University of Jerusalem, Israel}
\author{Julien de Rosny}
\affiliation{Institut Langevin, ESPCI Paris, PSL University, CNRS, France}
\author{Sébastien M. Popoff}
\email{Corresponding author: sebastien.popoff@espci.psl.eu}
\affiliation{Institut Langevin, ESPCI Paris, PSL University, CNRS, France}

\begin{abstract} 
Multimode optical fibers (MMFs) have gained renewed interest in the past decade,
emerging as a way to boost optical communication data-rates 
in the context of an expected saturation of current single-mode fiber-based networks. 
They are also attractive for endoscopic applications, 
offering the possibility to achieve a similar information content as multicore fibers,
but with a much smaller footprint, 
thus reducing the invasiveness of endoscopic procedures.
However, these advances are hindered by the unavoidable presence of disorder that affects 
the propagation of light in MMFs and
limits their practical applications.
We introduce here a general framework to study and avoid the effect of disorder 
in wave-based systems, and demonstrate its application for mutimode fibers. 
We experimentally find an almost complete set of optical channels that are resilient to disorder induced by strong deformations. 
These \textit{deformation principal modes} are obtained by only 
exploiting measurements for weak perturbations 
harnessing the generalized Wigner-Smith operator. 
We explain this effect by 
demonstrating that, 
even for a high level of disorder,
the propagation of light in MMFs can 
be characterized by just a few key properties.
These results are made possible thanks to a precise and fast 
estimation of the modal transmission matrix of the fiber 
which relies on a model-based optimization using 
deep learning frameworks. 
\end{abstract}

\maketitle

\section{Introduction}

The description of light transport in Multimode Fibers (MMFs) has been widely studied since the '70s, 
with a complete analytical understanding available in the case of an ideal straight fiber~\cite{marcuse1974theory}.
However, imperfections of the fabrication, geometrical deformations, or changes of the environmental conditions 
introduce randomness that drastically modifies their transmission properties. 
When light injected in one mode statistically explores all the other modes with the same probability, 
i.e. in the \textit{strong coupling regime},
some average properties can be predicted~\cite{Ho2011}.
However, from a few centimeters to a few kilometers, 
typical MMF systems are neither in the no coupling nor in the strong coupling regime; 
disorder strongly influences light propagation but some aspects of the ordered behavior survive~\cite{Gambling1975Mode, Ryf2012Mode, Ryf2015Mode}.
This intermediate regime has been little investigated so far due to the difficulty to experimentally characterize the effect of disorder on the modal content of the fibers. 
Understanding the transition between these two regimes remains an important challenge for
optical telecommunications, endoscopic imaging, and micromanipulation applications.

It is well known that injecting coherent light into an MMF results in the observation 
of a random pattern of bright and dark spots at the output, 
called \textit{speckle pattern}.
However, unlike scattering media, the observation of a speckle is not in itself a signature of disorder.
Indeed, perfect straight fibers also exhibit this property due to the existence of intermodal dispersion~\cite{ploshner2015seeing}.
As long as multiple modes are excited,
they quickly accumulate seemingly random relative phases leading to such complex interference patterns. 
In the past decade, the measurement of the Transmission Matrix (TM) emerged as a tool of choice to characterize and control the propagation of light in complex but deterministic optical linear systems.
Initially introduced in the context of scattering media~\cite{Popoff2010Measuring,Choi2011Overcoming,Popoff2011Image}, it consists in measuring the field-field linear relation between an input plane and an output plane.
This concept was then applied to MMFs, unlocking new applications for 
endoscopic imaging~\cite{Cizmar2012exploiting,choi2012scanner,papadopoulos2012focusing},
micromanipulation~\cite{Bianchi2012A},
quantum information processing~\cite{leedumrongwatthanakun2020programmable},
and for the control of optical channels for telecommunications~\cite{Carpenter2015Observation}.
It allowed in particular to demonstrate the robustness of the propagating modes in the case of short step-index fibers~\cite{ploshner2015seeing} and bent graded-index fibers~\cite{BoonzajerFlaes2018Robustness}.
However, the observation of the TM does not directly allow assessing the level of disorder, as dispersion and mode interference lead to the observation of a seemingly random matrix, even without disorder.
Only when represented in the basis of the propagating modes does the TM allow us to fully capture the spatial propagation properties of the MMF. 
This can be done by
directly injecting light and measuring the field in the mode basis~\cite{Carpenter2014_110}
or by numerically projecting a TM measured in a basis of diffraction-limited spots~\cite{ploshner2015seeing}.
In both cases, a good characterization is achieved only for the low order modes, 
as going into higher-order modes places increasingly demanding requirements 
on the beam quality and on the alignment~\cite{Stutzki2011High-speed}. 
A numerical post-treatment was demonstrated~\cite{ploshner2015seeing} 
to correct the TM measurement, but it still requires a careful and time-consuming procedure.
Moreover, such an approach assumes that there is little to no disorder in the fiber, 
which forbids the study of the transition from the weak to the strong mode coupling regimes. 
One of the main challenges of practical applications of multimode fibers 
is not only to understand the effect of disorder, 
but to avoid it altogether. 
In this context, the time-delay operator introduced in quantum
mechanics by E. Wigner and F. Smith~\cite{Wigner1955Lower,Smith1960Lifetime} 
has recently attracted renewed interest among the complex media  community.
For a lossless optical system characterized by its scattering matrix $\mathbf{S}$,
which links all input channels to all output ones,
the Wigner-Smith operator is constructed using the frequency derivative of $\mathbf{S}$, 
and defined as 
$
\mathbf{Q} = -i\mathbf{S}^{-1}\partial_\omega \mathbf{S}
$.
Interestingly, the eigenstates of this operator, 
also called \textit{principal modes},
are insensitive to small variations of the frequency.
The possibility to use wavefront shaping techniques to generate those input states
opened new applications to improve some properties of light transport,
such as to generate particle-like wave packets in chaotic cavities~\cite{Rotter2011Generating}
and in scattering media~\cite{Gerardin2016Particlelike,Bohm2018In},
or to optimize energy storage in scattering media~\cite{Durand2019Optimizing}.

For MMFs, the scattering matrix can be approximated by the TM, 
whose measurement gives access to the principal modes.
In the context of telecommunications, 
they are particularly attractive as they do not suffer from modal dispersion
to the first order~\cite{Fan2005Principal}.
Their ability to be stable over a large bandwidth was observed in the case of
weak~\cite{Carpenter2015Observation} and strong disorder~\cite{xiong2016principal}.
The possibility to find channels invariant to small modifications can be extended to other parameters than the frequency using the Generalized Wigner-Smith (GWS) operator~\cite{Ambichl_2017,Horodynski2020Optimal}. 
These studies focused on the interaction between waves and 
localized targets in scattering environments in the microwave regime.

In the present paper, we first introduce a new approach 
that relaxes most of the experimental constraints on the procedure
to measure quickly and accurately the TM of an MMF in the mode basis.
It uses numerical tools based on a modern machine learning framework. 
We demonstrate the ability to use the knowledge of the TM for small deformations to find
an almost complete set of channels using the GWS operator, 
the deformation principal modes,
that are insensitive to strong perturbations.
To understand this effect, we show that, all across the deformation range, the evolution of the TM 
can be characterized by only a few parameters that account for the 
mode coupling between close-by propagating modes.


\section{Misalignment and aberration robust calibration}

We first define the TM $\mathbf{H}_\text{pix}$ measured in the pixel basis of a modulator and a camera,
respectively located in planes conjugated with the input and output facets of a fiber.
Leveraging a fast digital micro-mirror modulator and InGaAs camera, 
we estimate $\mathbf{H}_\text{pix}$ with a 1~kHz frame rate in about 10 seconds. 
The principle of the experiment is presented in Fig.~\ref{fig:TM}a 
and detailed in Appendix~\ref{appendix:setup}.
Ideally, the mode basis representation of the TM can simply be recovered using:

\begin{equation}
\mathbf{H_{modes}} = \mathbf{M_o}^\dagger . \mathbf{H_{pix}}. \mathbf{M_i} \,\,,
\label{eq:proj}
\end{equation}

where $\mathbf{M_i}$ (resp. $\mathbf{M_o}$) represents the change of basis matrix between the input (resp. output) pixel basis and the mode basis of the fiber
(see Appendix~\ref{appendix:modes_theo} for the theoretical mode calculation). 
We use the orbital angular momentum modes basis, 
in which the modes are defined by a pair of indices $l$ and $m$, 
characterizing respectively to the radial oscillations and the orbital angular momentum. 
However, in the presence of slight aberrations or misalignments, 
the change of basis matrices cannot be inferred only from the calculation of the theoretical mode profiles of the fiber.
It leads to strong unwanted distortions of the mode basis TM, 
even with a carefully tuned setup~\cite{ploshner2015seeing}. 
Such an effect also occurs when working directly in the mode basis, 
as injected and detected modes, 
affected by the aberrations and misalignments,
are different from the true ones. 
To overcome this problem, we designed a numerical procedure based on the neural network framework PyTorch~\cite{pytorch2019},
taking advantage of Graphics Processing Units (GPUs) for optimized computational times.
Unlike neural networks, that consists of generalist layers, typically dense or convolutional layers, 
we use a model-based approach.
Our network is composed of custom layers, 
that each mimics the effect of an aberration by applying a Zernike phase polynomial to the change of basis matrix.
The structure of the network is represented in Fig.~\ref{fig:TM}d.
To take advantage of the deep learning optimization procedures,
based on gradient descent, 
each layer is differentiable.
Only one parameter per layer, 
the strength of the corresponding Zernike polynomial, 
has to be optimized.

We then train the model parameters to maximize $\norm{\mathbf{H}_\text{modes}}$, 
where $\norm{.}$ represents the $L_2$ norm (Frobenius norm) of a matrix.
Energy conservation imposes that the input and output projections performed in Eq.~\ref{eq:proj} lead to
$\norm{\mathbf{H}_\text{modes}} \leq \norm{\mathbf{H}_\text{pix}}$.
Since light can only be transmitted through the fiber by the propagating modes, 
these two quantities are equal when 
the matrices $\mathbf{M_i}$ and $\mathbf{M_o}$ correctly compensate for the aberrations
and misalignments.
Unlike neural networks, 
we do not need a large training set.
Indeed, we feed to the network only one experimentally measured matrix $\mathbf{H}_\text{pix}$.
Because of the low number of trainable parameters,  
46 corresponding to as much Zernike polynomials plus one for a global scaling on each facet,
the optimization takes only a few seconds to converge for a 110 mode fiber. 
More details about the numerical approach are provided in Appendix~\ref{appendix:model}. 
We effectively shifted the complexity of the acquisition 
from the experimental setup quality 
to the numerical optimization.
It allows changing the fiber in study in a matter of a few seconds,
without the need for a precise alignment procedure.

We show in Fig.~\ref{fig:TM}b 
the reconstructed mode basis TM
$\mathbf{H}_\text{modes}$ assuming no aberration
for a 30 cm unperturbed 110 mode graded-index fiber. 
The TM exhibits little symmetry with high losses for the highest order modes,
reflecting the effect of the aberrations and misalignments.
Moreover, only $49\%$  
of the energy is conserved in the mode basis.
After optimization (Fig.~\ref{fig:TM}c), about $94\%$ of the energy is conserved. 
Moreover, the matrix shows a strong diagonal, 
which traduces a weak mode-coupling effect.
$92\%$ of the energy is in the block diagonal,
representing the groups of degenerate modes (in green in Fig.~\ref{fig:TM}d). 
It is important to stress that the optimization process only maximizes the total energy in the mode basis, 
the observed strong diagonal appears naturally.
The procedure leads to accurate corrections regardless of the level of disorder 
(see Supplemental Material S5 for reconstruction comparison for different levels of disorder). 

\begin{figure*}[t]
\centering
\includegraphics[width=0.9\textwidth]{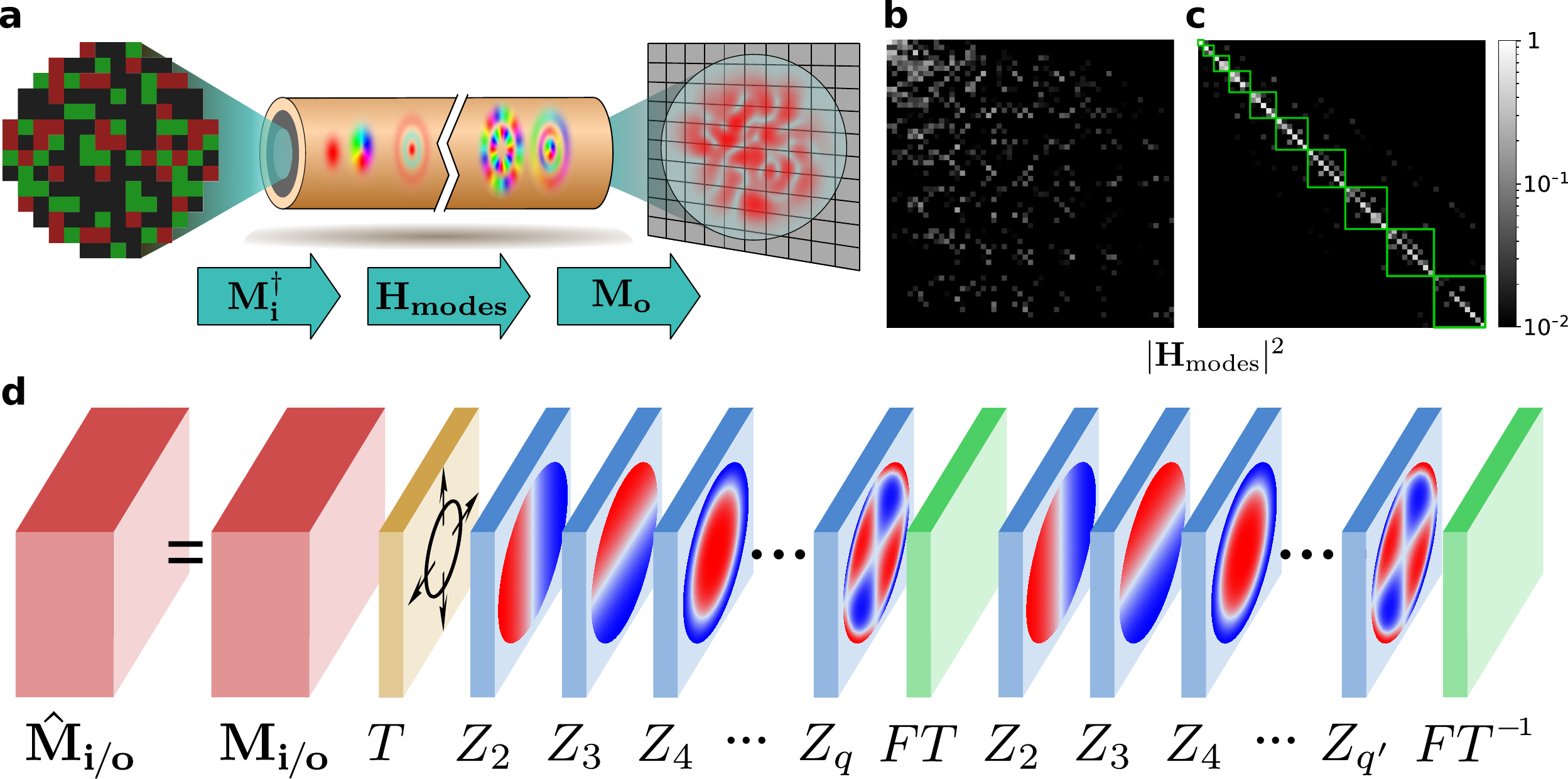}
\caption{
\textbf{Principle of the MMF TM reconstruction in the mode basis 
with the automatic compensation of the aberrations.}
\textbf{a}, Simplified sketch of the experiment: 
the input wavefront is modulated using a spatial light modulator and 
sent on the input facet of an MMF.
The light is transmitted through the propagating modes and the complex output field is imaged onto a camera. 
The TM is measured in the pixel basis and numerically projected onto the theoretical propagating modes. 
$\mathbf{M_i}$ and $\mathbf{M_o}$ represent to the input and output change of basis matrices, 
and $\mathbf{H_\text{modes}}$ is the TM in the mode basis.
\textbf{b}, \textbf{c}, Intensity of the experimentally measured TMs in the mode basis before and after the numerical compensation of the aberrations for one input and one output polarization. 
The green squares represent the groups of degenerate modes. 
The complete TMs are presented in Supplemental Material section S2.
\textbf{d}, Schematic of the model architecture used for the compensation of the aberrations.
$\mathbf{M_i}$ and $\mathbf{M_o}$ are modified by differentiable and trainable layers representing 
a homothety (yellow layer) 
and phase aberrations characterized by Zernike phase polynomials (blue layers).
The Fourier transforms (green layers) allow applying aberrations in the direct and
the Fourier planes.
The models for the two input and output mode conversions are trained 
simultaneously against a merit function that maximizes the energy in the projected matrix $\mathbf{H_\text{modes}}$ (see Appendix~\ref{appendix:model}).
}
\label{fig:TM}
\end{figure*}


\section{Perturbation insensitive channels}

\begin{figure*}[ht]
\includegraphics[width=0.95\textwidth]{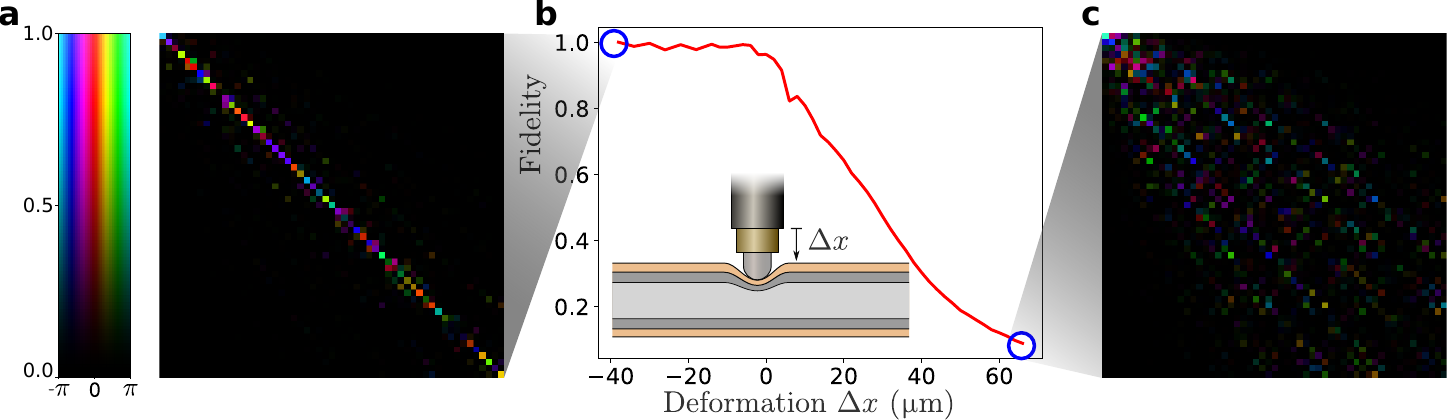}
\caption{
\textbf{Effect of deformations on the mode basis TM.}
\textbf{a}, \textbf{c}, Representation of $\mathbf{H}_\text{modes}$ 
for no deformation and when a transverse deformation ${\Delta x = 70}$~\textmu m is applied on the fiber.
\textbf{b} Fidelity between the TM of the deformed fiber 
and to the unperturbed one 
as a function of  $\Delta x$.
The fidelity is defined as
$F_c = Tr(|\mathbf{H}_\text{modes}\left(\Delta x\right) . 
\mathbf{H}_\text{modes}\left(\Delta x = 0\right)^ \dagger |^2)
/\sqrt{Tr\left(|\mathbf{H}_\text{modes}\left(\Delta x\right)|^2\right)\,
Tr\left(|\mathbf{H}_\text{modes}\left(\Delta x = 0\right)|^2\right)}$.
In inset, we represent a sketch of the deformation procedure.
}
\label{fig:deformation}
\end{figure*}

To learn how to be insensitive to disorder, 
we first characterize the full mode basis TM of an MMF 
when we introduce and gradually increase a perturbation.
We apply a controlled deformation on the fiber along an axis orthogonal
to the propagation direction (Fig.~\ref{fig:deformation}). 
Deformations of the fiber core leads to mode dependent losses, back-reflection, and mode coupling 
that hinders telecommunication applications. 
Qualitatively, strong deformations have the effect of progressively populating the off-diagonal elements of the TM 
while reducing the energy on the diagonal.
Thanks to the precise modal projection, 
we could observe for the first time the crossover from a nearly diagonal TM (weak coupling)~\cite{ploshner2015seeing}
to a seemingly random TM (strong coupling)~\cite{xiong2016principal}. 

\new{ 
	We first compute the correction for the TM of the unperturbed fiber using our aberration compensation approach. 
	We then apply the same correction parameters for the measurements obtained when the deformation is applied. 
	It ensures that the correction procedure does not compensate for some effects of the perturbation.
}
The fidelity between the matrix of the deformed fiber
and the reference matrix for the unperturbed configuration 
(see caption in Fig.~\ref{fig:deformation})
decreases quickly as the displacement $\Delta x$ increases. 
While the transmission properties are strongly altered for large deformations, 
our goal is to find a set of channels that are little affected by them.
In the present work, the parameter of interest is the induced displacement $\Delta x$, 
we then study the GWS operator defined as: 

\begin{equation}
\mathbf{Q}_{\Delta x} = -\frac{i}{2}\left[\mathbf{H}_\text{modes}^{-1} .
\partial_{\Delta x} \mathbf{H}_\text{modes} - 
\left(\mathbf{H}_\text{modes}^{-1} . \partial_{\Delta x} \mathbf{H}_\text{modes} \right)^\dagger \right] \,\,.
\label{eq:WS}
\end{equation}    

The second term appears due to the fact that $\mathbf{H}_\text{modes}$ is not 
unitary~\cite{private_cummun_AG}.
We estimate the GWS operator for a small deformation $\Delta x = 14$ \textmu m. 
The derivative is numerically estimated using the approximation:
\begin{equation} 
\partial_{\Delta x} \mathbf{H}_\text{modes} \approx \frac{\mathbf{H}_\text{modes}\left({\Delta x_0+\delta x}\right) 
- \mathbf{H}_\text{modes}\left({\Delta x_0-\delta x}\right) }{2 \delta x} \,\,.
\end{equation}
We chose $\delta x = 8$ \textmu m to mitigate the effect of noise that appears for smaller differences $\delta x$.

\begin{figure*}[ht]
\includegraphics[width=0.99\textwidth]{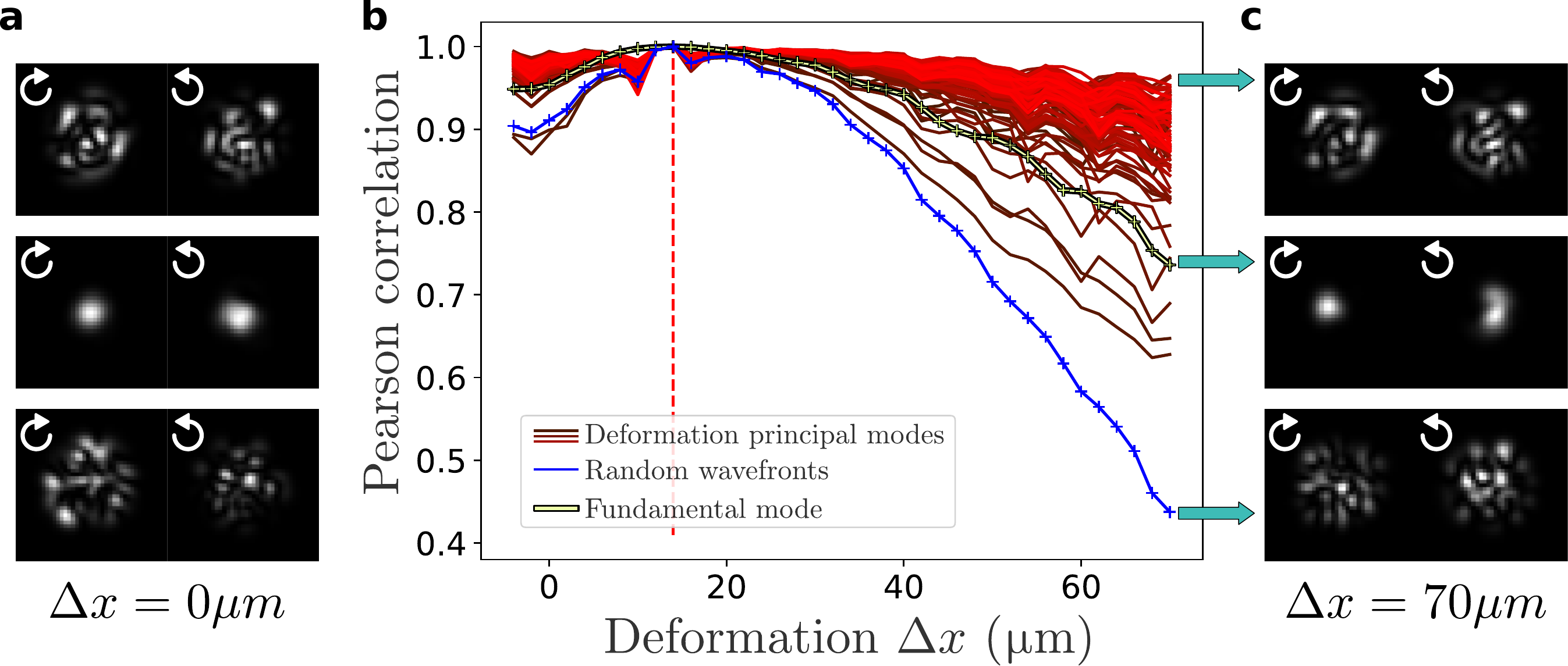}
\caption{
\textbf{Examination of the deformation principal modes.}
\textbf{a} and \textbf{c},
Output intensity profiles for the injection of 
a deformation principal mode, 
the fundamental mode of the fiber,
and a random wavefront for the minimal and maximal deformation. 
Each pair of patterns represents the intensity in the left and right circular polarization states of light.
\textbf{b}, Pearson correlation coefficient between 
the output intensity pattern at $\Delta x$
and the one at $\Delta x = 14$ \textmu m  
for all the deformation principal modes, 
the fundamental mode,
and after averaging over 20 random input wavefronts.
}
\label{fig:WS}
\end{figure*}

Its eigenmodes, referred to as the deformation principal modes,
are theoretically insensitive to the deformation parameter $\Delta x$ 
to the first order. 
We numerically compute their input profiles, and compare the output intensity patterns across the full range of deformations.
The stability of the deformation principal modes is shown in Fig.~\ref{fig:WS} 
as a function of $\Delta x$.
For comparison, we also test the injection of the fundamental mode, 
which is less affected by disorder than the other modes~\cite{Marcuse1976Field}, 
and random wavefronts (the correlation is averaged over 20 realizations). 
The best deformation principal modes keep a correlation above $95\%$ 
over the whole range of deformations
compared to the output profile for no deformation.
Moreover, all the principal modes but $4$ perform better than the fundamental mode. 
It is important to note that,
while the GWS operator is only estimated for small deformations, 
it provides an almost complete set of orthogonal channels robust to deformations, 
even for large values of $\Delta x$.
The fact that few generalized principal modes do not perform better than the fundamental mode
can be attributed to some of the fiber modes close to the cutoff being greatly attenuated (See Supplemental Material S5 for the singular value decomposition of the TM).

\section{Discussion}

To further investigate how the deformation principal modes, 
computed from the TMs for small deformations, 
can efficiently cancel the effect of large deformations,
we study the deformation matrix defined as: 

\begin{equation}
\mathbf{D}_j = \mathbf{H}_\text{modes}^{-1}(\Delta x = 0) 
. \mathbf{H}_\text{modes}(\Delta x_j) - \mathbb{I} \,\,.
\end{equation}

This matrix quantifies how $\mathbf{H}_\text{modes}\left(\Delta x\right)$  deviates from $\mathbf{H}_\text{modes}\left(\Delta x =0\right)$. 
It is equal to $0$ if the TM remains unchanged.
We want to determine the main characteristics 
that best describe how the TM is modified when the perturbation is applied. 
We then define a \textit{deformation operator} $\bar{\mathbf{D}}$
that links 
each value of the deformation $\Delta x_j$ 
to the corresponding deformation matrix $\mathbf{D}_j$. 
It characterizes the full evolution of the deformation of the matrix 
over the range of deformations applied. 
We compute the singular value decomposition of the operator $\bar{\mathbf{D}}$, 
it amounts to performing a principal component analysis. 
The deformation plays the same role as the different realizations in standard principal component analysis implementations 
(see Appendix~\ref{appendix:deformation}). 
We show in Fig.~\ref{fig:SVapprox}a 
that the first two singular values amount to more than $96\%$ of the total energy of the operator.
The corresponding singular components are represented in Fig.~\ref{fig:SVapprox}b 
for one  pair of input and output polarizations. 
As the deformation operator is computed for the whole range of deformations, 
the first principal components characterize the most important modifications applied 
to the TM during the deformation. 
It has been shown that,
for low perturbations introduced by thermal fluctuations,
the distortion of the TM can be parametrized by only one parameter~\cite{yammine2018time}.
To test here if the TM of a fiber under strong deformations 
can be  parametrized by only a few parameters,
we approximate the transmission matrix using just the first two components $\mathbf{U}_1$ and $\mathbf{U}_2$ using:

\begin{equation}
\mathbf{\hat{D}}_j = \alpha_j \mathbf{U}_1 +  \beta_j \mathbf{U}_2 \,\,.
\label{eq:sv_estimation}
\end{equation}

where $\alpha_j$ are $\beta_j$ are directly extracted from the singular value decomposition %
(see Appendix~\ref{appendix:deformation}). 
We show in Fig.~\ref{fig:SVapprox}c the fidelity between the estimated matrix 
${\mathbf{\hat{H}}_\text{modes}(\Delta x_j)} = {\mathbf{H}_\text{modes}(\Delta x = 0) .\left[ \mathbf{\hat{D}}_j + \mathbb{I} \right]}$ 
and the measured one. 
Surprisingly, all across the range of deformations, 
the TM can be estimated using only two parameters with a fidelity above $93\%$.
We can give a qualitative interpretation of the two significant components. 
$\mathbf{U}_1$ is close to identity, 
traducing the loss of energy in the diagonal compared to the reference TM at $\Delta x = 0$.
It is equivalent to the decay of the ballistic light in the presence of a scattering environment in free space.
The second vector $\mathbf{U}_2$ shows a well defined symmetric pattern that corresponds to an energy conversion 
between modes with close-by radial and angular momenta $l$ and $m$ (see Fig~\ref{fig:SVapprox}b, d).
This is consistent with the previous observations of mode coupling~\cite{Li2020compressively} 
 in bent graded index fiber. 
It corresponds to photons being injected in one mode and leaving the fiber in a close-by mode, 
which corresponds to photons whose direction has been modified once by the perturbation. 
This phenomenon is analogous to the conversion between ballistic and single scattered photons in scattering media. 
This physical interpretation is made possible thanks to the precise correction of the aberrations 
that would otherwise destroy the symmetries of $\mathbf{H}_\text{modes}$.

The fact that the TM can be estimated precisely using only 
two terms, only one of them accounting for mode coupling, 
is counter-intuitive considering the fact that $\mathbf{H}_\text{modes}$ 
shows a seemingly random aspect for high order modes at large deformations (see Fig.~\ref{fig:deformation}c).
Coupling between modes further away in the $l$ and $m$ space can occur, 
it is the equivalent of multiple scattered photons in scattering media.
However, strong mode coupling also comes 
with important losses due to coupling to non-guided modes that leak out of the fiber~\cite{Marcuse1976Field}, 
leading to a low energy contribution of this effect.
The fact that the same component $\mathbf{U}_2$ dominates the mode coupling effect 
for the whole deformation range explains how the deformation principal modes,
estimated for low deformations,
are still valid for strong deformations.

\begin{figure*}[ht]
\includegraphics[width=0.95\textwidth]{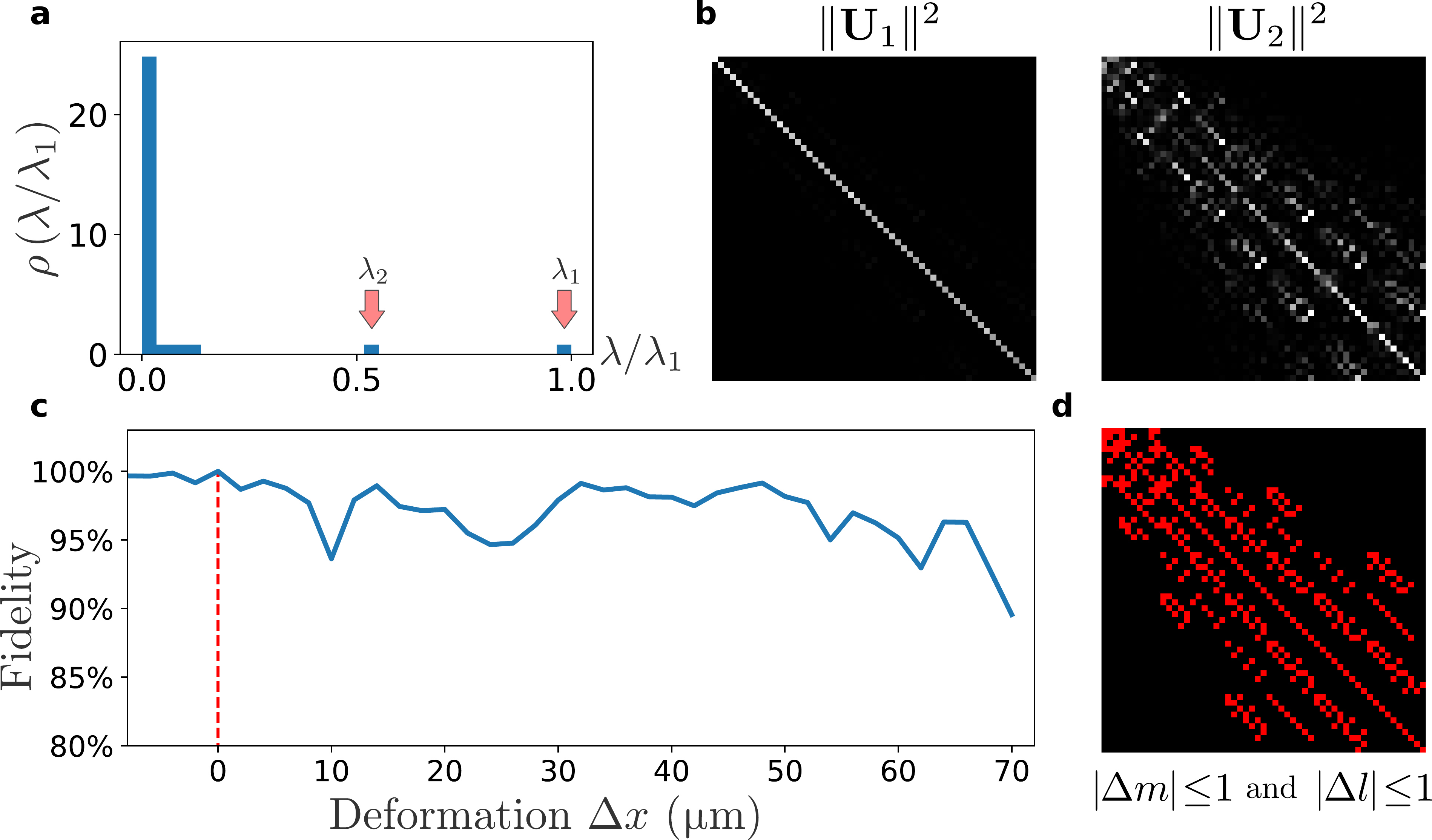}
\caption{
\textbf{Analysis of the effect of deformations.}
\textbf{a}, Singular value distribution of the deformation operator representing the full range of deformations. 
\textbf{b}, Intensity of the first two singular components $\mathbf{U_1}$ and $\mathbf{U_2}$.
\textbf{d}, Matrix representing the coupling between close-by modes,  
i.e. with a radial order $l$ difference equal or lower than $1$
and an orbital momentum $m$ difference equal or lower than $1$.
\textbf{c}, Fidelity between the measured TM in the mode basis 
and the approximated one using Eq.~\ref{eq:sv_estimation} (blue). 
The fidelity is defined similarly as in the caption of Fig.~\ref{fig:deformation}b.}

\label{fig:SVapprox}
\end{figure*}

\section{Conclusion}

To summarize, we present a framework to study the effect of disorder in MMFs, 
allowing, in a matter of seconds, to fully characterize 
light propagation in the mode basis. 
As precise predictions for the effect of perturbations on multimode fibers in real-life situations 
is currently lacking,
our approach provides a way to quantify such perturbations using measured transmission matrices  
and could serve as a benchmark for the study theoretical models. 
We harness this approach to observe for the first time the existence of deformation principal modes,  
that are robust against strong deformations, and 
that can be found by only using the knowledge of the fiber properties for small deformations.
This can be explained by the predominance in the transmission properties 
of the coupling between nearby modes,
even for large deformations.
We emphasize that our framework is general and can be used to study 
any linear propagation system  
regardless of the presence or the type of perturbations.
Moreover, as 
the complexity of handling the effect of the aberrations and misalignments 
in the TM estimation 
is rejected onto a fast automatic post-processing, 
our approach is virtually robust to any optical system imperfections, 
allowing \textit{plug-and-play} operations suitable for real-life applications.

\section*{Acknowledgements}
\noindent The authors kindly thank Arthur Goetschy 
and Esben R. Andresen 
for fruitful discussions.
M.W.M, J.R. and S.M.P acknowledge the French \textit{Agence Nationale pour la Recherche} (grant No. ANR-16-CE25-0008-01 MOLOTOF 
and grant No. ANR-20-CE24-0016 MUPHTA) and the  Labex WIFI (ANR-10-LABX-24, ANR-10-IDEX-0001-02 PSL*). 
Y.B. is supported by the Zuckerman STEM Leadership Program. 
Y.B. and S.M.P acknowledge the France-Israel grant (PRC1672) supported by
the Israel's Ministry of Science
and Technology and the France's
\textit{Centre National de la Recherche Scientifique} (CNRS).

\section*{Data and code availability}

\noindent Raw and processed data,
custom modules,
and sample codes for pre- and post-processing are  available in the dedicated repository~\cite{repo}.

\appendix
\section{\uppercase{Experimental setup}}
\label{appendix:setup}

\begin{figure}[ht]
  \includegraphics[width=0.85\textwidth]{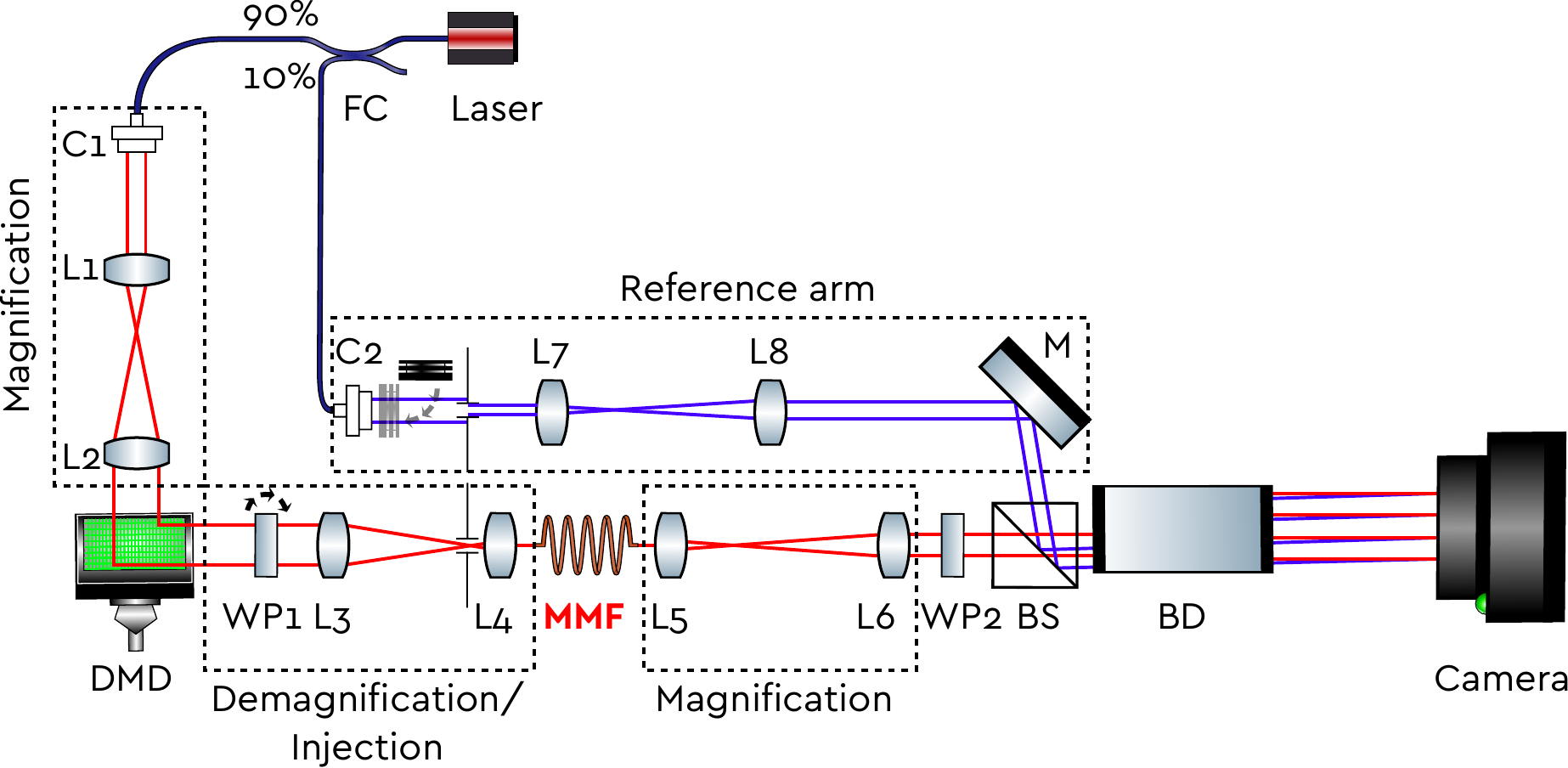}
  \caption{\textbf{Experimental setup.}
  L$_\text{i}$: lenses,
  C$_\text{i}$: collimators,
  WP$_\text{i}$: quarter-wave plates,
  M: mirror,
  D: diaphragm,
  BD: beam displacer.
  }
  \label{fig:setup}
\end{figure}

The optical setup is represented in Fig.~\ref{fig:setup}. 
The light source consists of a continuous linearly polarized laser beam at 1550 nm (TeraXion NLL)
injected into a 10:90 polarization-maintaining fiber coupler (PNH1550R2F1).
The 90$\%$ arm is collimated and expanded to illuminate a DMD (Vialux V-650L) composed of 1280 by 800 pixels with a pitch of $10.8$ \textmu m 
working at a maximum frame rate of 10.7 kHz. 
The light is converted into the left or right circular polarization using a quarter-wave plate and a motorized precision rotation mount (PRM1/MZ8). 
Two lenses allow the conjugation of the DMD plane with the surface of a standard 30 cm OM2 ($50$ \textmu m core) graded-index multimode fiber, consisting of a glass core, a glass cladding and an acrylate coating.
The MMF input facet is held by a fiber connector (Thorlabs B30128C3) and a bare fiber terminator (Thorlabs BFT1), 
and is mounted onto a 5-axis translation stage (Thorlabs APY001/M and MAX311D/M).
The output facet is positionned into a V-groove (HFV002) and held by magnetic clamps. 
The fiber is maintained approximately straight. 
Roughly at half the length, we place a V-groove to support the fiber 
where we introduce a deformation.
The perturbation is applied on the fiber by pressing on it using a $50 $ nm precision DC servo motor actuator (Thorlabs Z812).
Magnetic clamps are placed on both sides of the servo motor to prevent the fiber from slipping when the deformation increases. 
The coating absorbs a significant part of the deformation 
of the fiber, 
so that the deformation applied to the fiber core is proportional 
but smaller than the translation value $\Delta x$.
The output facet of the fiber is imaged onto an InGaAs camera (Xenics Cheetah 640-CL 400Hz) through a beam displacer (Thorlabs BD40)
that spatially separates the two polarization contributions in two different areas of the pixel array.
A reduced region of interest allows achieving frame rates of about 1 kHz.
The 10$\%$ arm of the fiber beam splitter is used to illuminate the camera with a tilted reference arm in an off-axis configuration~\cite{Cuche2000Spatial}.

\section{\uppercase{Measurement of the pixel basis TM}}
\label{appendix:TM}

The modulation of the input field is achieved using the Lee hologram method~\cite{Lee1978Computer}.
It allows performing complex amplitude modulation using a binary amplitude  DMD~\cite{Conkey2012High}.
The input modulation patterns consist in square layouts
of $ N^\text{in}_\text{pix} = 35 \times 35 = 1225$ square macropixels of size $20 \times 20$ pixels. 
We imprint on each macropixel a periodic pattern 
of bright (modulation state \textit{on}) 
and dark (modulation state \textit{off}) stripes.
Each macropixel effectively acts as a small grating. 
The first order of diffraction is selected with a diaphragm represented 
in Fig.~\ref{fig:setup}. 
In the plane of the input facet of the MMF,
the modulation on the macropixels can be 
switched \textit{on} or \textit{off} by removing the periodic pattern. 
The phase of the pixels can be modified by offsetting the grating patterns 
on each macropixel. 
We use a grating period of two pixels, giving access to only
two levels of phase modulation, $0$ and $\pi$. 
We can then create three complex amplitude states; $0$, $1$ and $-1$. 
Sequences of patterns are generated and sent to the control board of the DMD where they are stored in the on-board memory. 
The sequence is then displayed at a 1kHz frame-rate on the DMD, 
which triggers the acquisition of the frames on the camera.

A tutorial on the Lee holograms is made available on our website: 
\cite{WFS}. 
This modulation procedure has been implemented in the Python module \texttt{SLMlayout}~\cite{popoff2020pSLMlayout} 
and the interface control of the DMD was done using the Python module \texttt{ALP4lib}~\cite{popoff2020pALP4lib}. 
We developed, share, and maintain both packages.\\

The complex output field is measured using an off-axis holographic technique~\cite{Cuche2000Spatial}.
We share a tutorial on the off-axis holography and some sample codes 
on our website: \cite{WFS}. 
The complex field is simultaneously measured for the two orthogonal 
circular polarization states.
A quarter-wave plate converts the left and right circular polarizations 
into two linear orthogonal polarizations. 
A beam splitter combines the reference arm and the signal arm, 
and a beam-displacer projects the contributions 
from the two polarization states on two different regions 
of the camera. 
For each polarization, the optical field is projected onto 
a square pattern of $41 \times 41$ square macropixels. 
The field is averaged over each macropixel. 
The output field for each input wavefront
is encoded into a vector of size 
$2 N_\text{pix}^\text{out} = 2 \times 41 \times 41 = 3362$, 
where $N_\text{pix}^\text{out}$ is the number of macropixels 
for each polarization.\\

The first step of our experiment is to estimate experimentally the TM
in the pixel basis $\mathbf{H}_\text{pix}$. 
This matrix describes the linear relationship between the field on one pixel of the modulator 
to the field on one pixel of the camera. 
We send a set of input wavefronts described by the vectors $X_i$, $i\in \left[1\twodots N_\text{masks}\right]$, that represent the field on all the input macropixels. 
The corresponding output field patterns are represented 
in the basis of the camera macropixels 
by the vectors 
$Y_i$,  $i\in \left[1\twodots N_\text{masks}\right]$. 
The relation between the input and output fields reads:

\begin{equation}
    Y_i = \mathbf{H}_\text{pix} . X_i\quad \forall i\in \left[1\twodots N_\text{masks}\right]
    \,\,.
    \label{eq:TM_pix}
\end{equation}

Let's call $\mathbf{X}$ (resp. $\mathbf{Y}$) the matrix that represents the stack of vectors $X_i$ (resp.  $Y_i$).
Eq.~\ref{eq:TM_pix} can be rewritten:

\begin{equation}
    \mathbf{Y} = \mathbf{H}_\text{pix} . \mathbf{X} \,\,.
\end{equation}

An estimation $\mathbf{\hat{H}}_\text{pix}$ of the TM can be found by 
using each vector of the the canonical basis for the input excitation patterns, 
i.e. using $\mathbf{X} = \mathbb{I}$. 
It gives direct access to the TM using 
$\hat{\mathbf{H}}_\text{pix} = \mathbf{Y}$.
One can also use any orthogonal basis, 
such as the Hadamard basis that is convenient for phase-only modulation~\cite{Popoff2010Measuring}, 
so that $\hat{\mathbf{H}}_\text{pix} = \mathbf{Y}. \mathbf{X}^{-1}$. 
However, in the presence of noise, 
or if one or more measurements fail, 
the quality of the reconstructed matrix is significantly altered. 
To mitigate those effects, we chose to use a set of random vectors $X_i$ 
with $N_\text{masks} > N^\text{in}_\text{pix}$.
We can then estimate the TM using:

\begin{equation}
    \hat{\mathbf{H}}_\text{pix} = \mathbf{Y} . \mathbf{X}^+ \,\,,
\end{equation}

where $.^+$ represents the Moore-Penrose pseudo-inverse.
We chose $X_i$ to be random patterns where the modulation on each pixel can take the value $0$, $-1$ or $1$.
For each pattern, the percentage of \textit{off} pixels, 
i.e. taking the value $0$,
is drawn from a uniform distribution 
between $60\%$ and $80\%$. 
The percentage of \textit{on} pixels taking the value $1$ 
is drawn from a uniform distribution between $40\%$ and $60\%$, 
the other pixels taking the value $-1$.
The positions of the pixels are random.
We chose $N_\text{masks} = 6 \times N^\text{in}_\text{pix} = 7350 $ to ensure the existence and the stability of the pseudo-inverse of $\mathbf{X}$.\\

By changing the input polarization state, 
we measure separately the two corresponding sub-matrices.
They are finally combined into a large matrix 
of size $2 N^\text{out}_\text{pix} \times 2 N^\text{in}_\text{pix}$.

\section{\uppercase{Singular value decomposition of the deformation operator}}
\label{appendix:deformation}

We first reshape the stack of the matrices $\mathbf{D}_j$ 
as a 2-dimensional matrix
$\mathbf{\bar{D}}$ 
of size ${N_{\text{modes}}^2 \times N_{\Delta x}}$,
where $N_{\Delta x}$ is the number of deformations, and
 $N_{\text{modes}} = 110$ the number of propagating modes.
It links each deformation, indexed by $j$, 
to all the elements of the matrix  $\mathbf{D}_j$, 
indexed by the composite index $\{kl\} \in \left[1\twodots N_{\text{modes}}^2\right]$. 
The range $\left[1\twodots N_{\Delta x}\right]$ of the index $j$ corresponds to deformations 
between $\Delta x = 0$~\textmu m and the maximal deformation $\Delta x = 70$~\textmu m.
Next, we calculate the singular value decomposition of this operator:

\begin{equation}
\mathbf{\bar{D}} = \mathbf{U} . \mathbf{\Lambda} . \mathbf{V}^\dagger \,\,.
\label{eq:SVD}
\end{equation}

$\mathbf{\Lambda}$ is a diagonal matrix 
of size $N_{\Delta x} \times N_{\Delta x}$
containing the singular values, 
whose distribution is represented in Fig.~\ref{fig:SVapprox}a.
 $\mathbf{U}$ is a matrix containing the corresponding output singular vectors 
$U_i$, $i \in \left[1 \twodots N_{\Delta x}\right]$. 
They can be reshaped as 2-dimensional matrices $\mathbf{U}_i$ of size $N_{\text{modes}} \times N_{\text{modes}}$. 
For any given deformation,
we can approximate $\mathbf{D}_j$ using only $\mathbf{U}_1$ and $\mathbf{U}_2$ with Eq.~\ref{eq:sv_estimation}.
It amounts to replacing $\mathbf{\Lambda}$ in Eq.~\ref{eq:SVD} by $\tilde{\mathbf{\Lambda}}$ defined by $\tilde{\Lambda}_{11} = \Lambda_{11} = \lambda_1$, $\tilde{\Lambda}_{22} = \Lambda_{22} = \lambda_2$ and $\tilde{\Lambda}_{kl} = 0$ for all other values.

The coefficient $\alpha_j$ and $\beta_j$ in Eq.~\ref{eq:sv_estimation} are then expressed by:

\begin{align}
\alpha_j = \lambda_{1} V_{1j}^*\,\,,\\
\beta_j = \lambda_{2} V_{2j}^*\,\,.
\end{align}

\begin{figure}[ht]
  \includegraphics[width=0.75\textwidth]{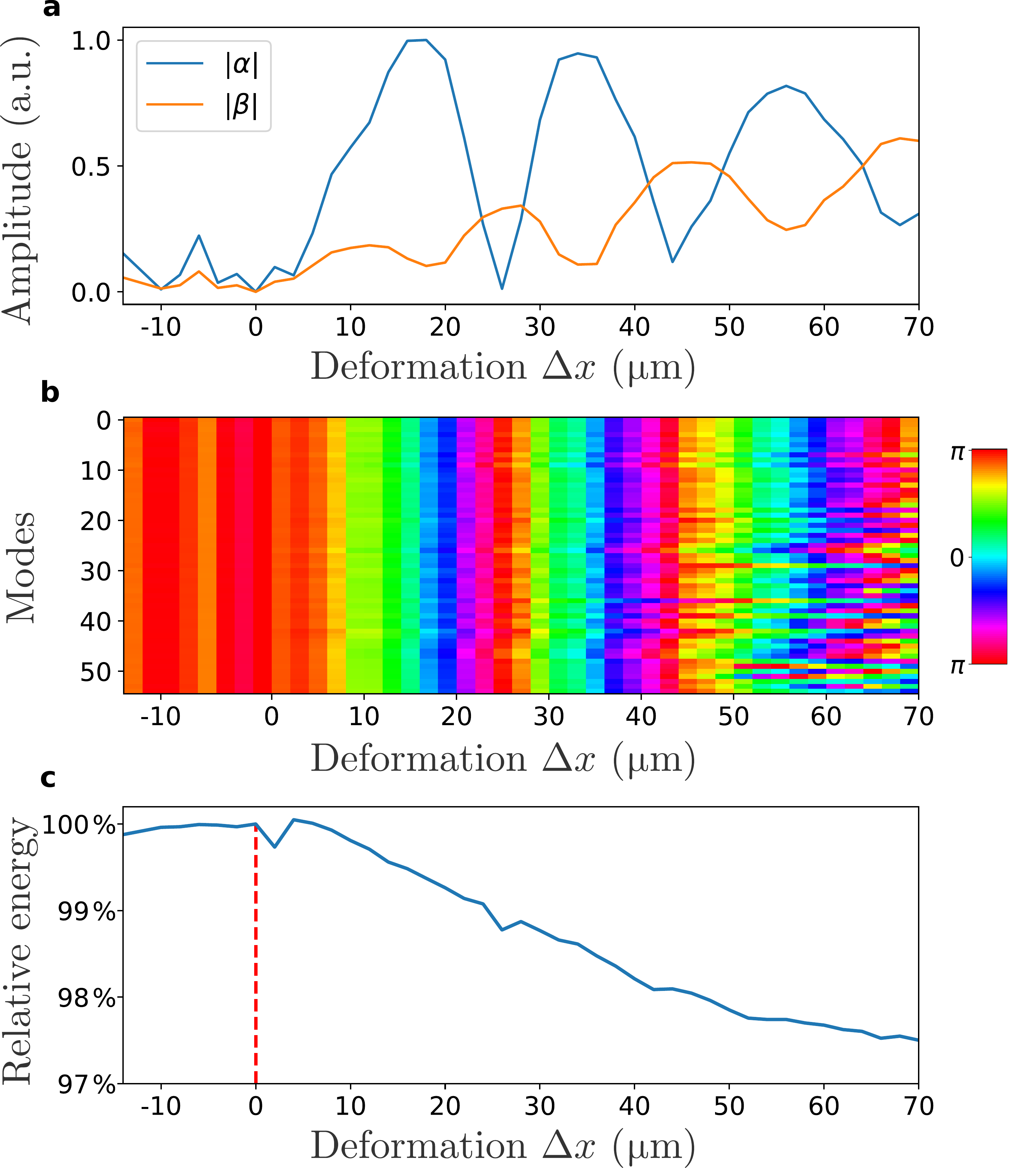}
  \caption{
  \textbf{Analysis of the evolution of the transmission matrix.}
  \textbf{a}, Coefficients of $\alpha$ and $\beta$ in Eq.~\ref{eq:sv_estimation} in arbitrary units as a function of the deformation. Coefficients are normalized by the maximal value of $\left|\alpha\right|$.
  \textbf{b}, Relative phase between the diagonal elements of the mode basis TM as a function of the deformation and the same diagonal elements for $\Delta x = 0$. 
  \textbf{c}, Energy variation of the mode basis TM as a function of the deformation evaluated as
  $\norm{\mathbf{H}_\text{modes}\left(\Delta x\right)}^2/\norm{\mathbf{H}_\text{modes}\left(\Delta x =0\right)}^2$
  }
  \label{fig:coeffs}
\end{figure}

We represent in Fig.~\ref{fig:coeffs}a the evolution of the absolute value of the coefficients of $\alpha$ and $\beta$ as a function of the deformation. 
The contribution of $\alpha$ is dominant for small deformations and globally decreases as the deformation increases. 
Conversely, the contribution of $\beta$ is small for small deformation and globally increases with the deformation. 
This trend confirms that the first effect to appear is the loss of energy on the diagonal, 
due to the effect of $\mathbf{U_1}$,
and then the coupling to neighboring modes in the momentum space, 
due to the effect of $\mathbf{U_2}$. 
We observe that this global trend is modulated by a periodic oscillation. 
The beating between the two contributions can be attributed to the fact that
the two physical effects are not fully decoupled in the two operators 
as $\mathbf{U_2}$ also has significant energy on the diagonal, 
that modifies the energy of the ballistic photon similarly to $\mathbf{U_1}$. 
It has been shown that, in addition to mode coupling, 
deformations are associated with a global rescaling of the fiber which induces phase shifts that dominate for small deformations~\cite{ploshner2015seeing, BoonzajerFlaes2018Robustness,yammine2018time}. 
We represent in  Fig.~\ref{fig:coeffs}b the evolution of the phase on the diagonal of the modes basis TM. 
We observe an oscillation with the same periodicity as the beating between $\alpha$ and $\beta$.
As the perturbation increases, one expects higher-order coupling effects to become significant in Eq~\ref{eq:sv_estimation}, 
which would be associated with the coupling between modes further away in the momentum space. 
However, such an effect increases the chance for the photons to couple to non-guided modes, leading to losses~\cite{Marcuse1976Field}. 
We show in Fig.~\ref{fig:coeffs}c the variation of the energy of the mode basis TM as a function of the deformation. 
Losses increase with the deformation up to approximately 2.5$\%$, confirming that higher-order coupling effects are still weak in this regime. 
We restrict ourselves in this study to deformations in the elastic regime of the material, 
higher deformations leading to non-reversible perturbations and permanent damage of the fiber.

\section{\uppercase{Calculation of the theoretical modes}}
\label{appendix:modes_theo}

The starting point of the mode projection operation is to consider the ideal modes of the fiber.
We want to estimate the modes profiles of a perfect straight graded-index fiber under the scalar approximation.
Graded-index fiber mode profiles and dispersion relation
do not have a closed-form analytical expression.
However, approximate analytical expressions can be found, for instance,  using perturbation theory or a variational approach~\cite{sharma1992solutions}.
Arguably the most widely used approximation is the Wentzel-Kramers-Brillouin (WKB) approximation. 
It leads to an analytical dispersion relation when assuming an infinite quadratic 
spatial profile of the refractive index. 
While leading to accurate estimations of the propagation constants, 
it has a limited accuracy for the expression of the spatial mode profiles~\cite{Gedeon1974Comparison,Maksymiuk2016On}, 
especially for low radial numbers $l$.
Finite difference methods are easy to implement numerically, 
but the 2D discretization of the field leads to high memory requirement and computation time,
and could lead to inaccuracies for high order modes. 
Because we consider axiosymmetric index $n(r)$ profiles, we want to simplify the system to solve a 1D problem that only depends on the radial coordinate $r$, 
allowing us to increase the accuracy and decrease the computation time.\\

The 2D scalar Helmholtz equation for a propagating mode can be written in the cylindrical coordinate system as

\begin{equation}
\partial_r^2 \psi(r,\phi) + \frac{1}{r}\partial_r \psi(r,\phi)+ \frac{1}{r^2}\partial_\phi^2 \psi(r,\phi)
+ \left[ k_0^2 n^2(r) - \beta^2\right] = 0\,\,,
\label{eq:hel_radial}
\end{equation}

where $\psi$ is the optical field,  
$\phi$ is the azimuthal coordinate, 
$\beta$ is the propagation constant, 
and $k_0 = 2\pi/\lambda$ with $\lambda$ the wavelength.\\

Because the refractive index only depends on the radial coordinate $r$ for a perfect graded-index fiber, 
we can separate the variables $r$ and $\phi$.
We are then looking for the orbital angular momentum modes of the form:

\begin{equation}
\psi_{ml} = f_l(r) e^{i m \phi}\,\,.
\end{equation}

with $l$ the radial order and $m$ the azimuthal order, which also corresponds to the 
orbital angular momentum.
Injecting this expression in equation~\ref{eq:hel_radial} leads to the 1D equation

\begin{equation}
d_r^2 f_l(r) + \frac{1}{r}d_r f_l(r) + \left[ k_0^2 n^2(r) - \beta^2 - \frac{m^2}{r^2}\right]f_l(r) = 0
\label{eq:hel_oam}
\end{equation}

The singularity at $r = 0$ arising from the $\frac{1}{r}$ term makes  direct finite difference methods unstable.
We can use the transformation:

\begin{equation}
g_l(r) = \frac{1}{f_l(r)} d_r f_l(r),
\label{eq:transformation}
\end{equation}

and rewrite equation~\ref{eq:hel_oam} as a quadratic Ricatti equation~\cite{Tamil1991Finite}:

\begin{equation}
d_r g_l(r) + P(r) + Q(r)g_l(r) + g^2_l(r) = 0,
\end{equation}

where

\begin{align}
Q(r) &= \frac{1}{r},\\
P(r) &= k_0^2 n^2(r) - \beta^2 - \frac{m^2}{r^2}.
\end{align}
A finite difference approximation of such equation leads to the recursive expression~\cite{Tamil1991Finite,file2016numerical}:

\begin{equation}
1+h g_l^{n+1} = \frac{1}{1+hQ_n/2}\left(h^2 P_n-2+\frac{1-h Q_n/2}{1+h g_l^n}\right),
\end{equation}

where $g_l^{n} = g_l(r_n)$,
$Q_{n} = Q(r_n)$,
$P_{n} = P(r_n)$,
and $h = r_{n+1} - r_n$ is the step size.\\

The expression \ref{eq:transformation} can then be discretized as:

\begin{equation}
f_l^{n+1} = f_l^{n} \left( 1+h g_l^{n}\right).
\end{equation}

To find the first steps to initialize the iteration, we need to consider the boundary conditions
at the center of the fiber core:

\begin{align}
\left.\frac{df}{dr}\right|_{r=0} = 0 & \quad \text{for } m = 0,\\
f(r=0) = 0 & \quad \text{for } m \neq 0.
\end{align}

For $m=0$, we discretize the functions at $r_n = n h-1/2$,
and initialize the functions with $f_l^0 = 1$ and $g_l^0 = 0$.
For $m\neq0$, we discretize the the functions at $r_n = n h$,
and initialize the functions with $f_l^0 = 0$, $f_l^1 = h$ and $g_l^1 = (1-h^2P_1)/h$.
For a given value of $m$, 
the propagation constants $\beta_{ml}$ that satisfy the Helmholtz equation, 
corresponding to the propagating modes, 
are the ones for which the field vanishes at large values of $r$.\\

The steps to find the modes of the fiber are the following:
We start with $m=0$, and perform a coarse scan of the propagation constant values 
between $\beta_{min} = k_0 n_{min}$ and $\beta_{max} = k_0 n_{max}$.
We choose $r_N > a$ large enough to assume that the field at this point, 
and thus $f_N$, should be vanishingly small.
The number of times $f_N(\beta)$ changes sign gives us the number of propagation modes for the current value of $m$. 
It corresponds to the maximal radial number $l$ admissible for the azimuthal number $m$.
We then use a binary search algorithm to find,
at a minimum computational cost,
the accurate admissible values of $\beta$ for each $l$, 
i.e. the values that minimize $f_N$ under a given tolerance value.
We then increment the value of $m$, and repeat the procedure.
We stop when no solution is found for the current value of $m$.\\

This procedure has been implemented in the Python module \texttt{pyMMF}~\cite{popoff2020pyMMF}
that we developed and share. 
Sample codes to compute the ideal modes of the MMF  are available at 
the dedicated repository~\cite{repo}. 

\section{\uppercase{Model-based optimization for the compensation 
of the aberrations and misalignements}}
\label{appendix:model}


Recent attempts were made to tackle the problem of modal decomposition 
using deep learning frameworks. 
As they used model-free neural network models,  
using standard convolutional~\cite{An2019cnn} 
or dense layers~\cite{Rothe2020dense}, 
these systems require large training sets
and significant amounts of memory.
Moreover, computational times and limited accuracy 
forbid their use for more than 10 modes 
(the training for 10 modes took about 43 hours with a $>$ 300,000 image training set in~\cite{Rothe2020dense}).
We developed here a model-based approach that only learns a few relevant parameters, is fast (a few seconds) to converge, and only requires one TM measurement. 
The  general principle is to apply to the change of basis matrices $\mathbf{M}_i$ and $\mathbf{M}_o$ a set of transformations that mimics the effect of aberrations and misalignments to compensate for the experiment's imperfections.
The schematic of the model is presented in Fig.1d of the main text. \\

In order to implement our model, we first need to use complex-valued matrix operations.
However, complex tensors are not natively supported by the PyTorch framework we use. 
To do so, we add a dimension to our data structure of size 2 
to encode the real part and the imaginary part of the complex values.
We then create a set of elementary operations: 
complex conjugation, 
element-wise, 
and matrix multiplications. 
The key parts of our approach are the layers 
that mimic the effect of aberrations represented by Zernike polynomials.
The input of each layer is a batch of complex 2D images of size 
$N_\text{modes} \times N_\text{pix} \times N_\text{pix} \times 2$. 
The effect of a layer $Z_k$, 
corresponding to the $k$-th Zernike polynomial, 
is to add, to each 2D image, a phase contribution. 
It amounts to transforming each input image $K_{ij}$, 
$(i,j) \in \left[1 \twodots N_\text{pix}\right] \times 
\left[1 \twodots N_\text{pix}\right]$ 
into a modified one $K'_{ij}$ using:

\begin{equation}
    K'_{ij} = K_{ij} e^{j \alpha_k F_k(r_{ij},\phi_{ij})} \,\,,
\end{equation}

where $F_k(r,\phi)$ is the $k$-th Zernike polynomial, 
$r_{ij}$ and $\phi_{ij}$ are the polar coordinates 
corresponding to the pixel indexed by $i$ and $j$,
and $\alpha_k$ is the weight of the aberration.
$\alpha_k$ is the only trainable parameter of the layer. 
The layer automatically calculates and stores the derivative 
of the output tensor with respect to this parameter, 
as required for the training process (backpropagation). 
By adding multiple Zernike layers, we simulate the effect of a high level of aberration. 
We perform a Fourier transform in the spatial dimensions 
and add other Zernike layers to simulate aberrations in the Fourier plane
(see Fig.1d of the main text). 
The first Zernike polynomials correspond to phase slopes in the $x$ and $y$ directions and to a parabolic phase. 
When applied in the Fourier plane, 
they introduce spatial shifts in the $x$ and $y$ directions and a defocus. 
It allows compensating for misalignments in the $x$, $y$ and $z$ directions.
Finally, we add a transformation $T$ that applies a global scaling transformation in the spatial dimensions.  
The scaling parameter is the only trainable parameter of this layer.\\

We treat separately each combination 
of input and output polarizations. 
For each optimization, we train simultaneously two models, 
one for the input and one for the output 
change of basis matrix. 
The input data corresponds to the matrices
$\mathbf{M}_i$ and $\mathbf{M}_o$ of respective size
$N^\text{modes} \times N^\text{in}_\text{pix}$ and
$N^\text{modes} \times N^\text{out}_\text{pix}$ 
that we compute using the approach detailed in the previous section.
We convert and reshape them as PyTorch tensors of sizes 
$
N^\text{modes} 
    \times N^\text{in}_\text{x} 
    \times N^\text{in}_\text{y}
    \times 2
$
and 
$
N^\text{modes} 
    \times N^\text{out}_\text{x} 
    \times N^\text{out}_\text{y}
    \times 2
$,
with $N^\text{modes}=55$ the number of modes per polarization, 
$ N^\text{in}_\text{x} = N^\text{in}_\text{y} 
= \sqrt{N^\text{in}_\text{pix}} = 35 $, 
and 
$ N^\text{out}_\text{x} = N^\text{out}_\text{y} 
= \sqrt{N^\text{out}_\text{pix}} = 41 $.
The first dimension is treated as the batch size in conventional neural networks.
The two models return new input and output change of basis matrices $\mathbf{M'}_i$ and $\mathbf{M'}_o$
that are used as input and output projectors on the pixel basis TM:

\begin{equation}
\mathbf{\hat{H}}_\text{modes} = \mathbf{M'_o}^\dagger . \mathbf{H}_\text{pix} . \mathbf{M'_i} \,\,,
\end{equation}

As explained in the main text, 
we know that an ideal compensation of the aberrations corresponds to maximizing $\norm{\mathbf{H}_\text{modes}}$, 
with $\norm{.}$ representing the $L_2$ norm (Frobenius norm) of a matrix. 
We choose as the cost function to minimize:

\begin{equation}
    \mathcal{L} = \frac{\norm{\mathbf{H}_{pix}}}{\norm{\mathbf{\hat{H}}_\text{modes}}}\,\,,
\end{equation}
where $\mathbf{H_{pix}}$ is the experimentally measured pixel basis TM.
\\

Finally, 
we run an optimizer based on a stochastic gradient descent approach 
(Adam optimizer~\cite{kingma2014adam}) 
to find the set of parameters 
(weights of the Zernike polynomials 
and the global scaling factors in input and output)
that minimizes the cost function $\mathcal{L}$. 
Once the optimization finished, 
the obtained change of basis matrices can be used 
on any newly acquired pixel basis TM as long as the setup stays unchanged.
\new{
The full optimization takes 
18 seconds on a computer with an Nvidia GeForce 2080 Ti GPU and a Xeon Gold 6142 CPU,  
36 seconds on the same computer with CPU computations only, 
and 51 seconds on a regular laptop with  an Intel i7-8550U CPU and no GPU. 
The gain of the GPU computation is expected to increase drastically when the number of modes increases, 
taking advantage of tensor manipulation optimizations on GPUs.
}\\

The full model, the custom layers, and sample codes 
of aberration correction using experimental data 
are available at the dedicated repository~\cite{repo}. 

\end{document}


\title{Learning and avoiding disorder in multimode fibers\\
Supplementary Information}

\author{Maxime W. Matth{\`e}s}
\affiliation{Institut Langevin, ESPCI Paris, PSL University, CNRS, France}
\author{Yaron Bromberg}
\affiliation{Racah Institute of Physics, The Hebrew University of Jerusalem, Israel}
\author{Julien de Rosny}
\affiliation{Institut Langevin, ESPCI Paris, PSL University, CNRS, France}
\author{Sébastien M. Popoff}
\affiliation{Institut Langevin, ESPCI Paris, PSL University, CNRS, France}

\maketitle

\clearpage

\section{Full pixel basis transmission matrices}

\begin{figure}[ht]
  \centering
\vspace{-5em}
\end{figure}  
  \begin{figure}[ht]
  \centering
  \includegraphics[width=0.70\textwidth]{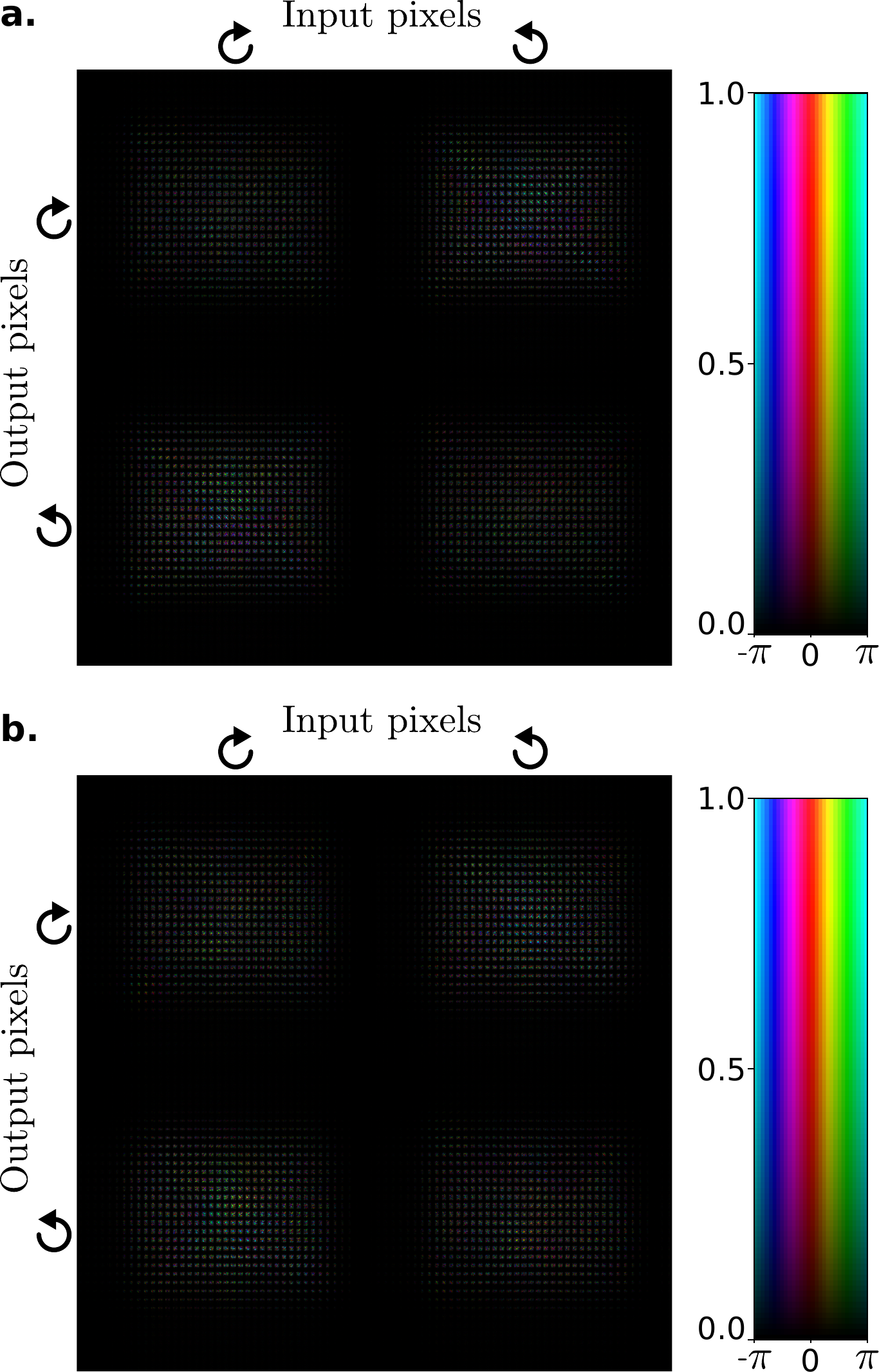}
  \caption{\textbf{Pixel basis TM.}
  Complex amplitude of the elements of the pixel basis TM for both left and right circular polarizations 
  in input and output 
  $\Delta x = 0$ \textmu m (\textbf{a.}) and 
  for $\Delta x = 70$ \textmu m (\textbf{b.}) 
  }
\end{figure}

\section{Full mode basis transmission matrices before correction}

\vspace{-1em}
\begin{figure}[H]
  \centering
  \includegraphics[width=0.7\textwidth]{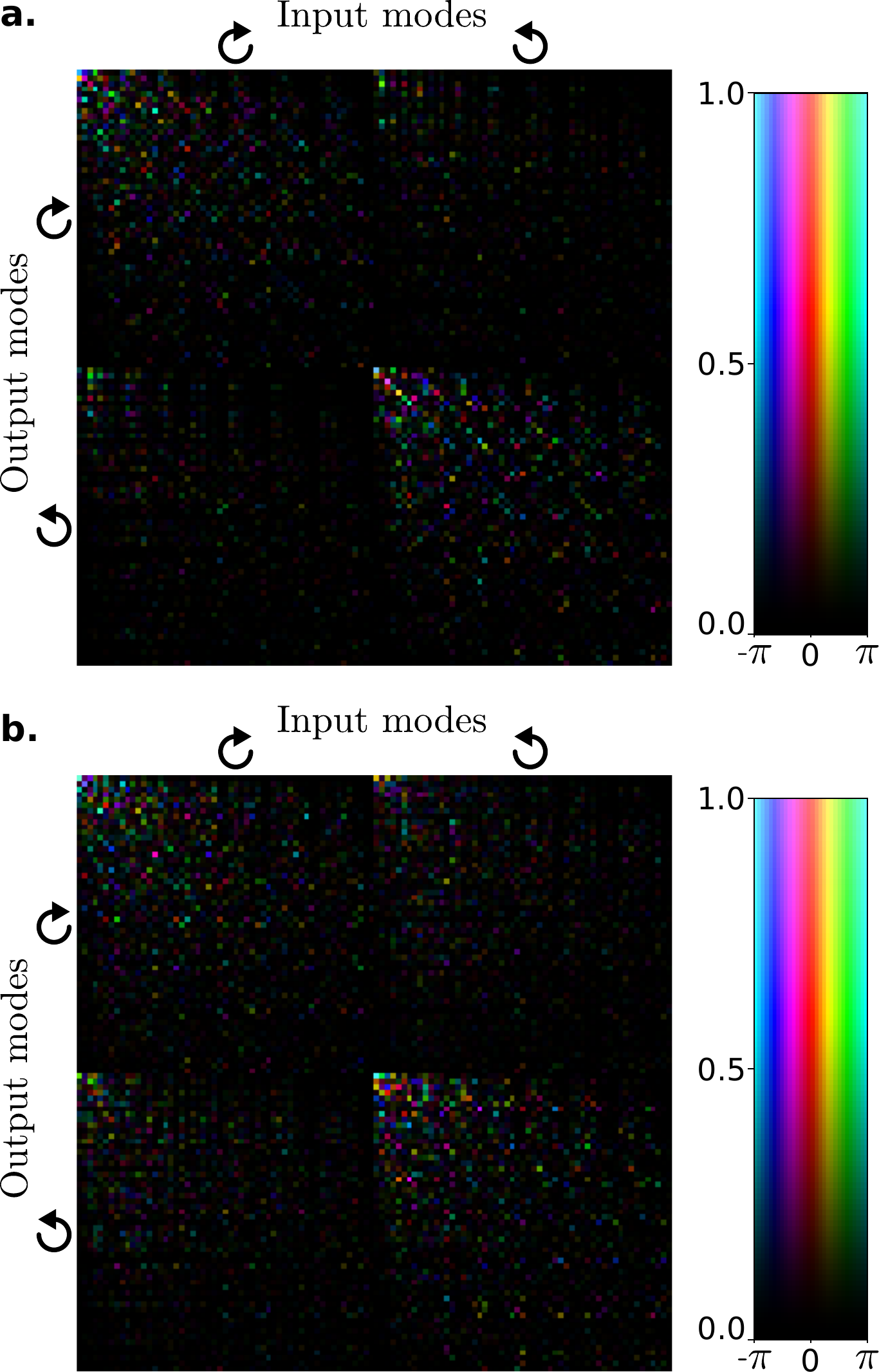}
  \caption{\textbf{Reference and perturbed mode basis TM before aberrations correction.}
  Complex amplitude of the elements of the TM projected on the mode basis
  before the compensation of the aberrations for $\Delta x = 0$ \textmu m (\textbf{a.}) and 
  for $\Delta x = 70$ \textmu m (\textbf{b.}) 
  for both left and right circular polarizations 
  in input and output.
  }
\end{figure}

\section{Full mode basis transmission matrices after correction}

\vspace{-1em}
\begin{figure}[H]
  \centering
  \includegraphics[width=0.70\textwidth]{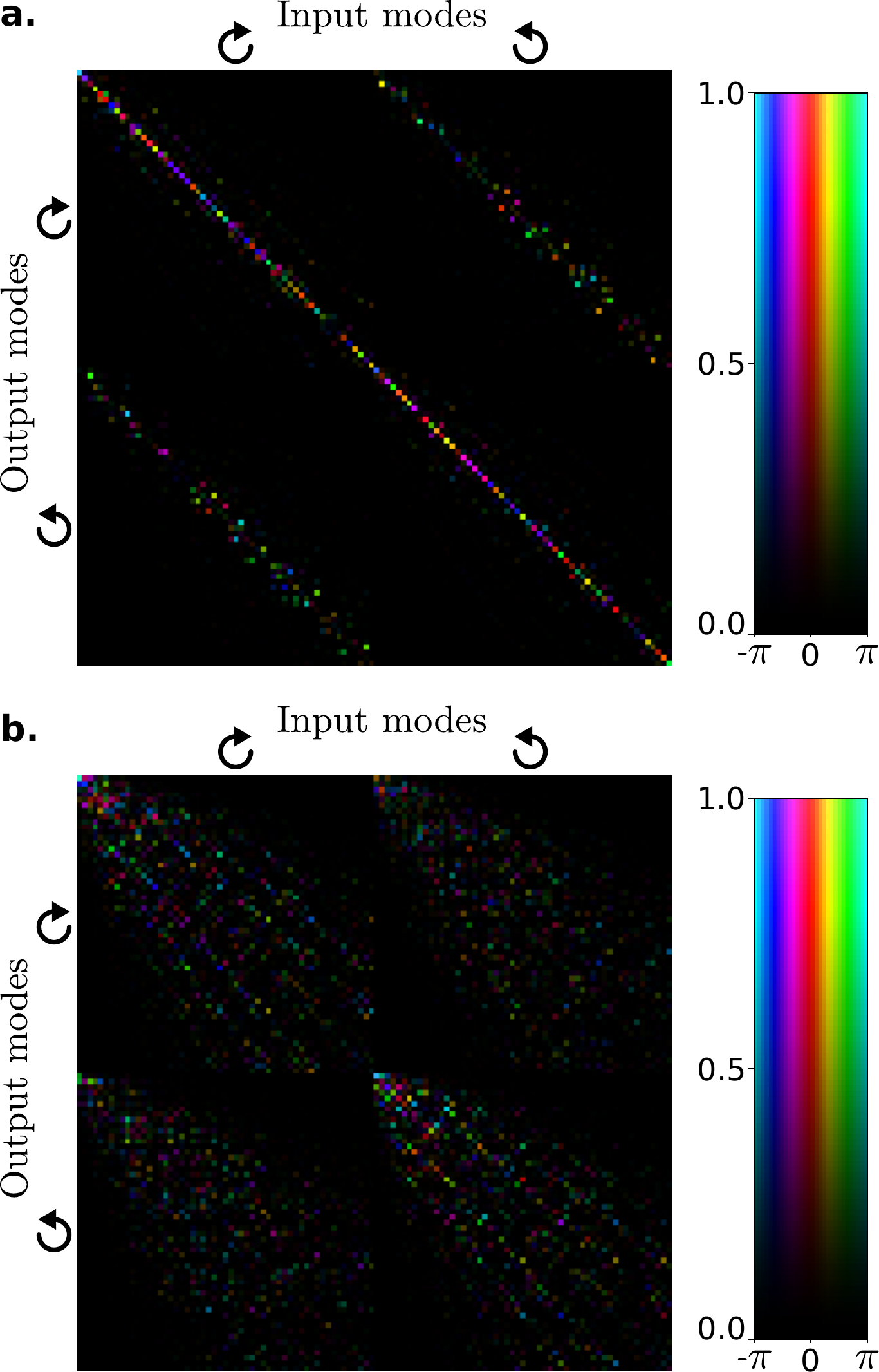}
  \caption{\textbf{Perturbed mode basis TM after aberrations correction.}
  Complex amplitude of the elements of the TM projected on the mode basis
  after the compensation of the aberrations for $\Delta x = 0$ \textmu m (\textbf{a.}) and 
  for $\Delta x = 70$ \textmu m (\textbf{b.}) 
  for both left and right circular polarizations 
  in input and output.
  }
\end{figure}

\section{Mode profiles before and after the correction of aberrations}

\begin{figure}[ht]
  \centering
  \includegraphics[width=0.75\textwidth]{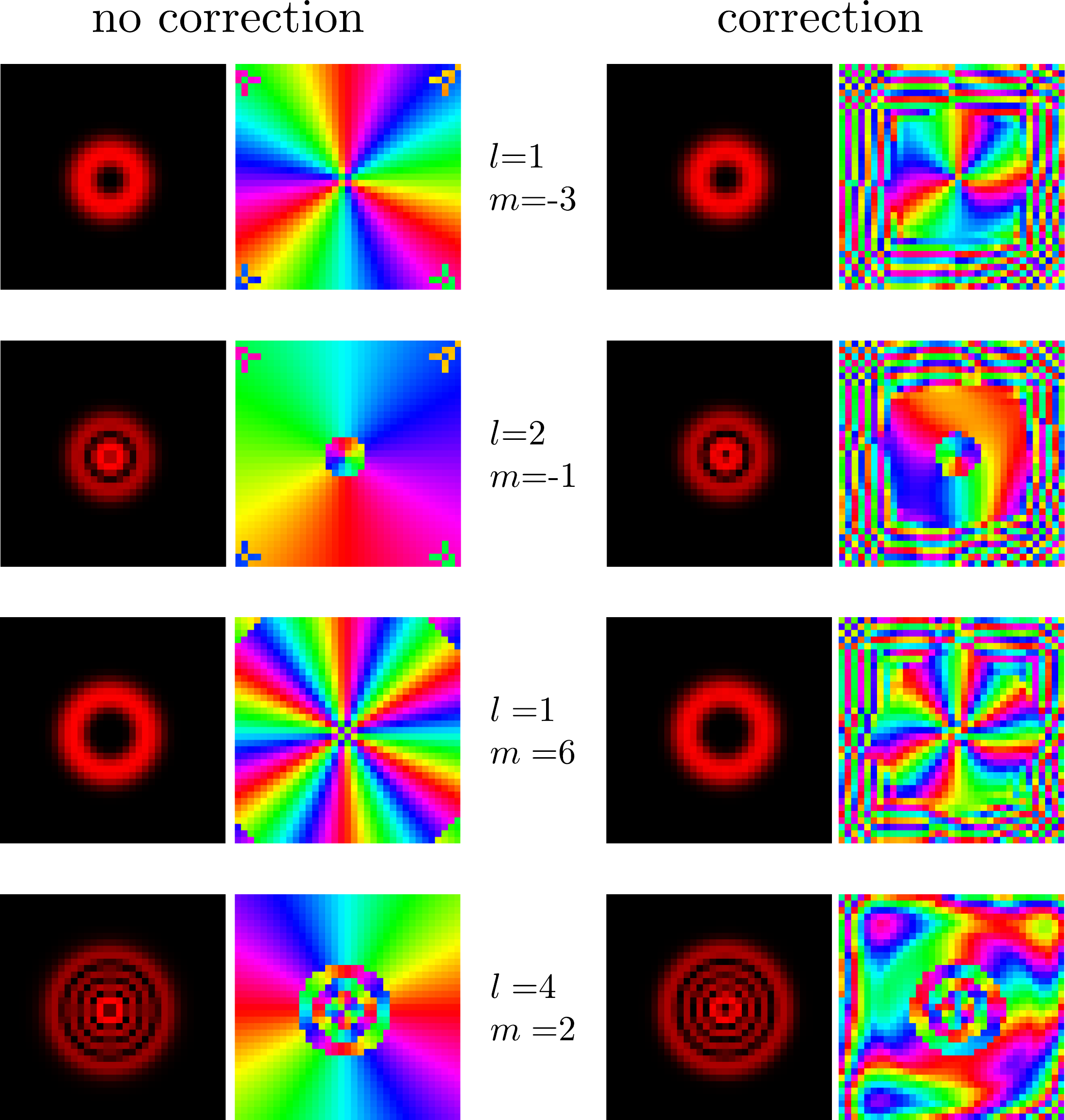}
  \caption{\textbf{Input mode intensity profiles.}
  Before (left) and after (right) the correction of aberrations
  for four selected fiber modes. 
  For each mode, we represent the amplitude (left) and the phase (right) of the optical field and indicate the radial and angular momenta $l$ and $m$.
  The corrections are computed for $\Delta x = 0$ \textmu m.  }

\end{figure}

\begin{figure}[H]
  \centering
  \includegraphics[width=0.75\textwidth]{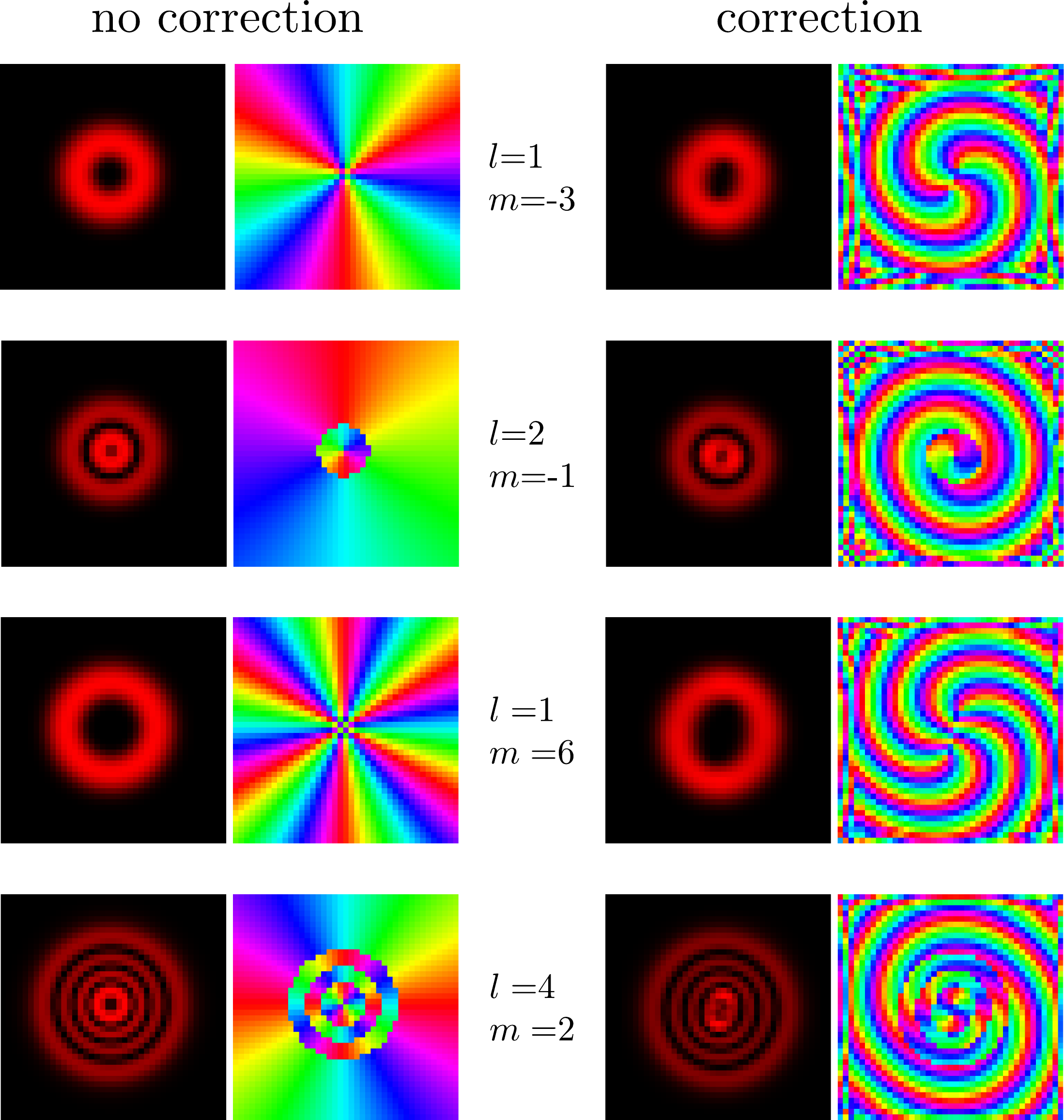}
  \caption{\textbf{Ouput mode intensity profiles.}
  Before (left) and after (right) the correction of aberrations
  for four selected fiber modes. 
  For each mode, we represent the amplitude (left) and the phase (right) of the optical field and indicate the radial and angular momenta $l$ and $m$.
  The corrections are computed for $\Delta x = 0$ \textmu m.
  }

\end{figure}

\section{Aberration correction with deformation}

\begin{figure}[H]
  \centering
  \includegraphics[width=0.6\textwidth]{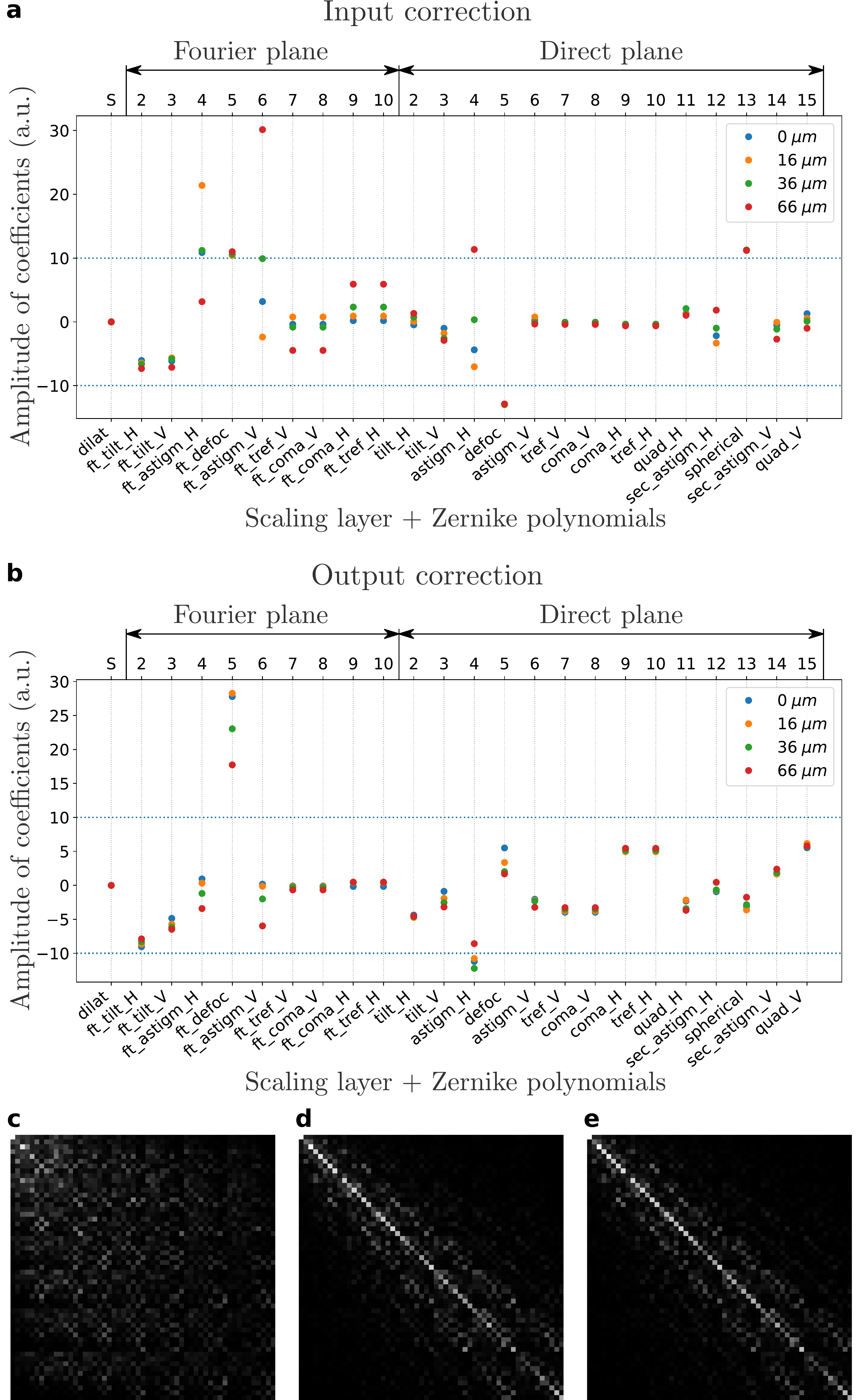}
  \caption{\textbf{Comparison of aberration correction for different perturbations.}
  \textbf{a}, and \textbf{b} Coefficients found by the optimization procedure for the scaling parameter (S) 
  and the Zernike polynomials (9 in the direct plane, 14 in the Fourier plane), 
  for different values of the deformation,
  corresponding to the input and output correction respectively.
  \textbf{c}, \textbf{d}, and \textbf{e}, Amplitude of the mode basis TM 
  for $\Delta x = 16$ \textmu m 
  without correction, 
  with the correction computed for $\Delta x = 16$ \textmu m, 
  and with the correction computed for $\Delta x = 0$ \textmu m.
  }
  \label{fig:corr_comp}
\end{figure}

We represent in Fig.~\ref{fig:corr_comp}a and b the coefficients found for the 
scaling parameter and the Zernike polynomials (9 in the direct plane, 14 in the Fourier plane) 
for different values of the deformation. 
The coefficients are similar except for some variations, 
mostly for the astigmatism aberrations in both planes. 
We attribute these fluctuations to the different degrees of freedom not being totally independent, 
hence different combinations could lead to the same transformation of the modes. 
To confirm that the resulting transformations are indeed similar, 
we compute the corrected mode basis TM for $\Delta x = 16$ \textmu m 
using the correction obtained for $\Delta x = 16$ \textmu m  (Fig.~\ref{fig:corr_comp}d) 
and using the correction obtained for $\Delta x = 0$ \textmu m  (Fig.~\ref{fig:corr_comp}e). 
The error between the TMs, 
defining the quadratic error between two matrices $\mathbf{A}$ and $\mathbf{B}$ 
as $Err = \norm{\mathbf{|A|}-\mathbf{|B|}}^2/\norm{\mathbf{|A|}}\norm{\mathbf{|B|}}$, 
is about 3 \%. 
The fidelity, defined as 
$F_c = Tr(|\mathbf{A}.\mathbf{B}^ \dagger |^2)
/\sqrt{Tr\left(|\mathbf{A}|^2\right)\,
Tr\left(|\mathbf{B}|^2\right)}$, 
is about 94 \%. 
The remaining difference between the two reconstructions 
can be attributed to actual changes in the physical system, 
such as a longitudinal displacement of the fiber 
when the pressure is applied. 
This explanation is supported by the variations in the defocus coefficient 
in the Fourier plane as=t the output of the fiber.

\section{Singular value decomposition}

\begin{figure}[H]
  \centering
  \includegraphics[width=0.60\textwidth]{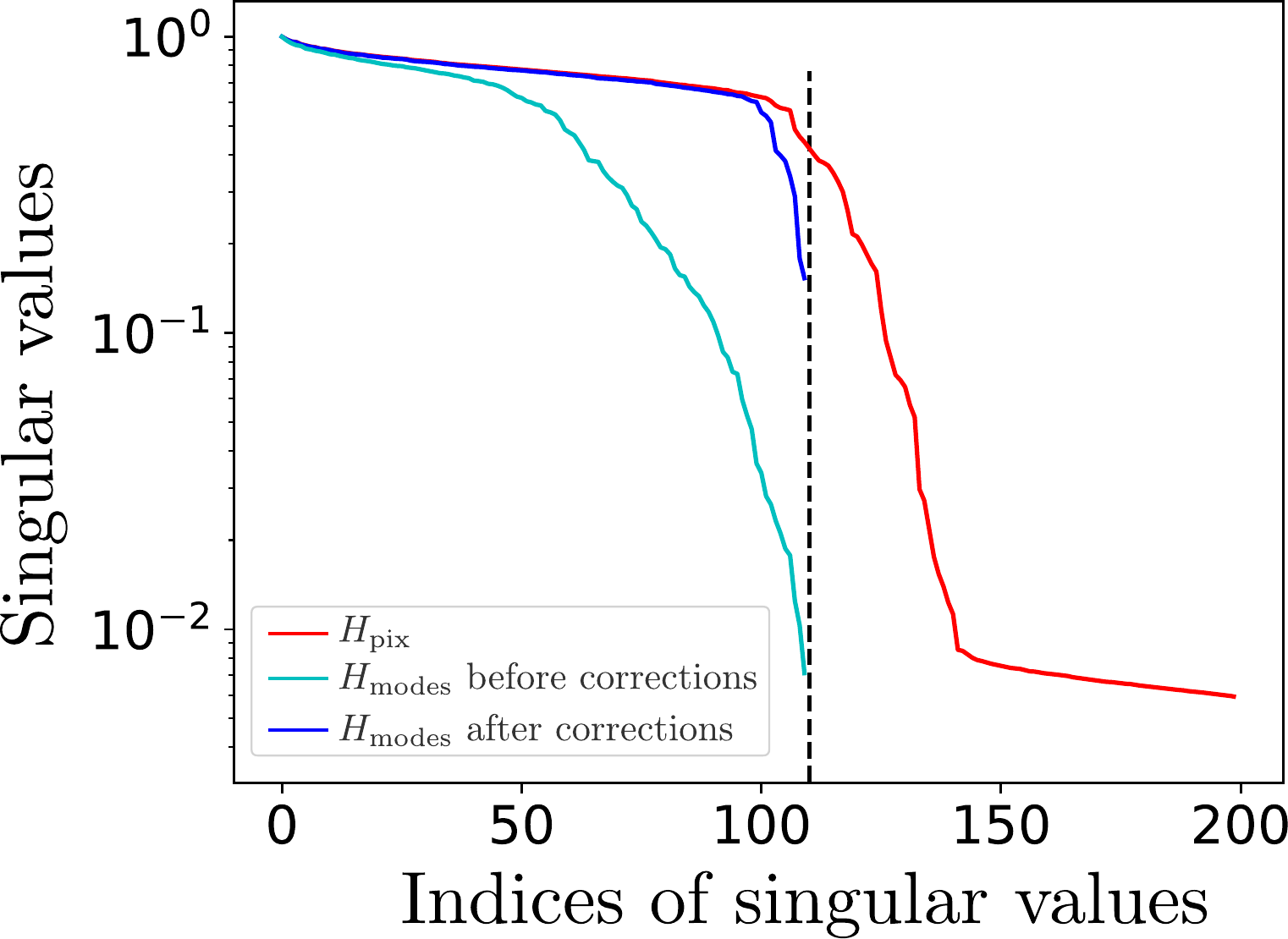}
  \caption{\textbf{Singular values of the reference TM.}
  Singular values of the pixel basis TM  (red), 
  of the TM projected onto the mode basis before the correction of the aberrations (cyan),
  and after the numerical optimization (blue) 
  for $\Delta x = 0$ \textmu m.
  The vertical dashed line indicates the theoretical number of modes (110).
  Singular values are normalized by the maximal one in each case.
  }
  
    \centering
  \includegraphics[width=0.60\textwidth]{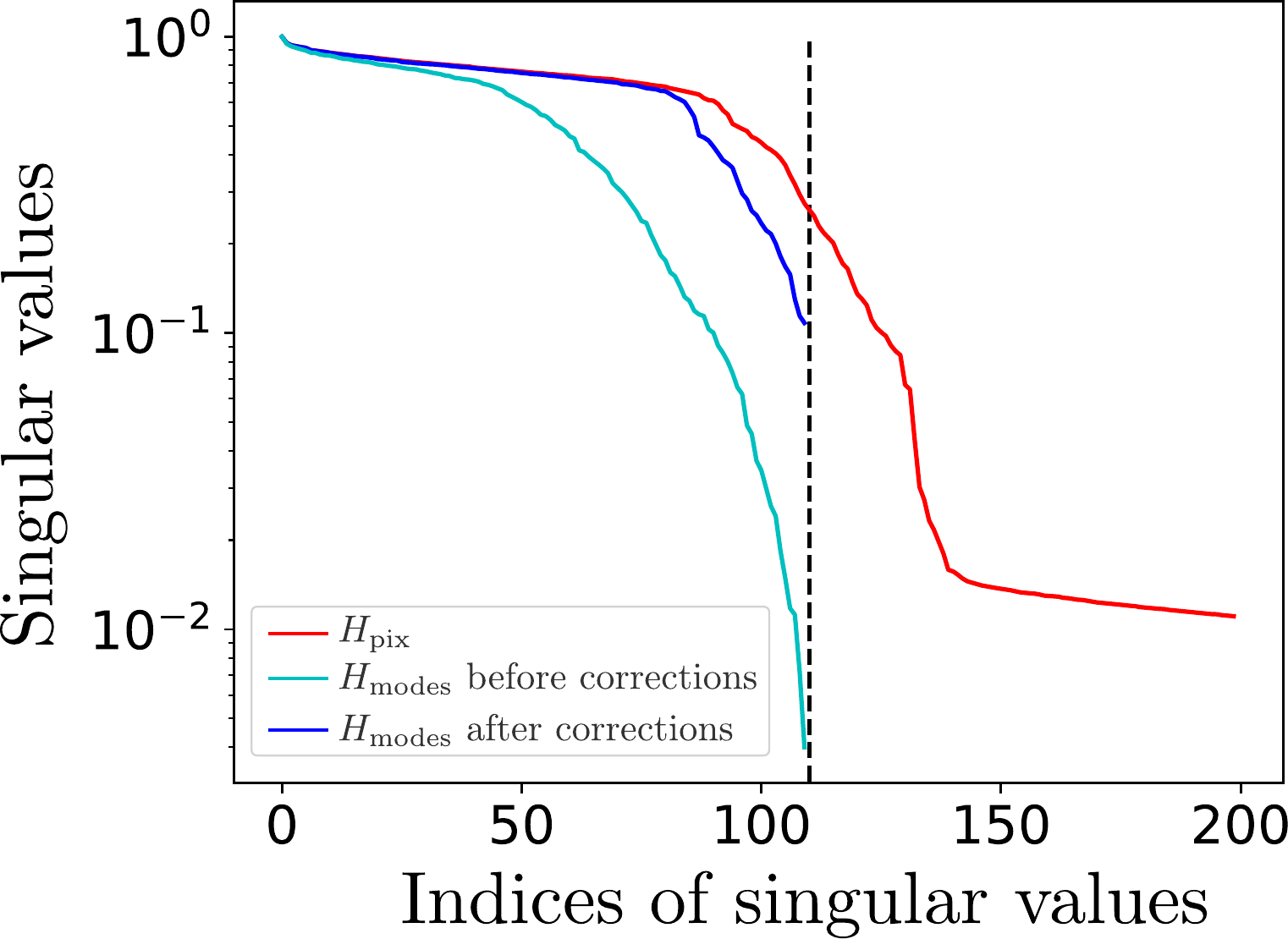}
  \caption{\textbf{Singular values of the perturbed TM.}
  Singular values of the pixel basis TM  (red), 
  of the TM projected onto the mode basis before the correction of the aberrations (cyan),
  and after the numerical optimization (blue)
   for $\Delta x = 70$ \textmu m.
  The vertical dashed line indicates the theoretical number of modes (110).
  Singular values are normalized by the maximal one in each case.
  }
\end{figure}
\clearpage

\section{Intensity profiles of a few deformation principal modes}

\begin{figure}[H]
  \centering
  \includegraphics[width=0.60\textwidth]{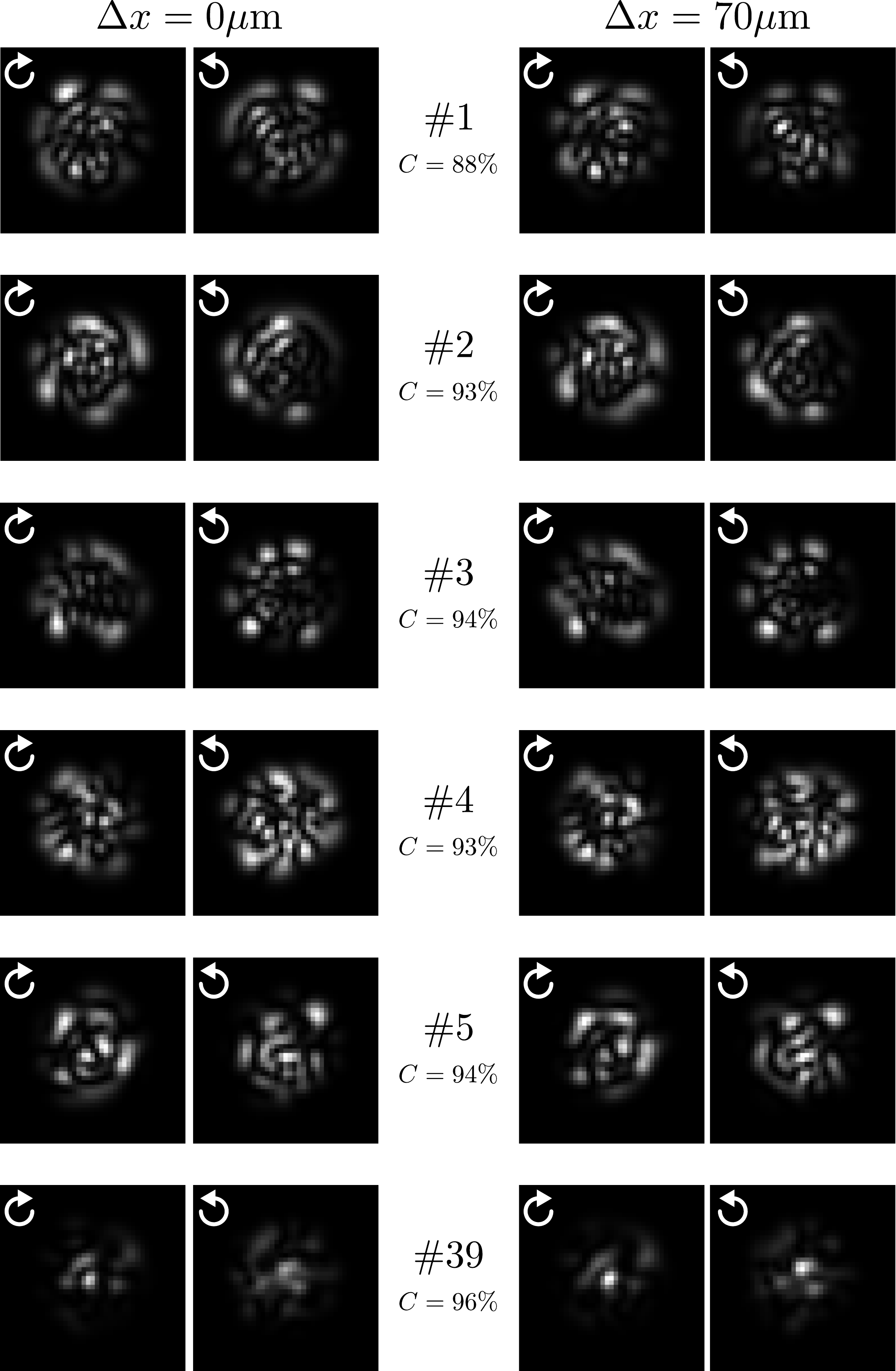}
  \caption{\textbf{Intensity profiles of a few deformation principal modes.}
  Intensity profiles for both polarizations 
  for the first 5 principal modes and the 39\textsuperscript{th} one, 
  which is the one that gives the best correlation between the output profiles at minimal and at maximal deformation.
  The output intensity profiles for $\Delta x = 0$~\textmu m
  are presented in the left column and 
  for ${\Delta x = 70}$~\textmu m in the right column. 
  The correlation between the intensity profiles at both deformation values are presented in inset.
  }
\end{figure}
\clearpage

\section{Projection of the deformation principal modes onto the fiber modes}

\begin{figure}[H]
  \centering
  \includegraphics[width=0.95\textwidth]{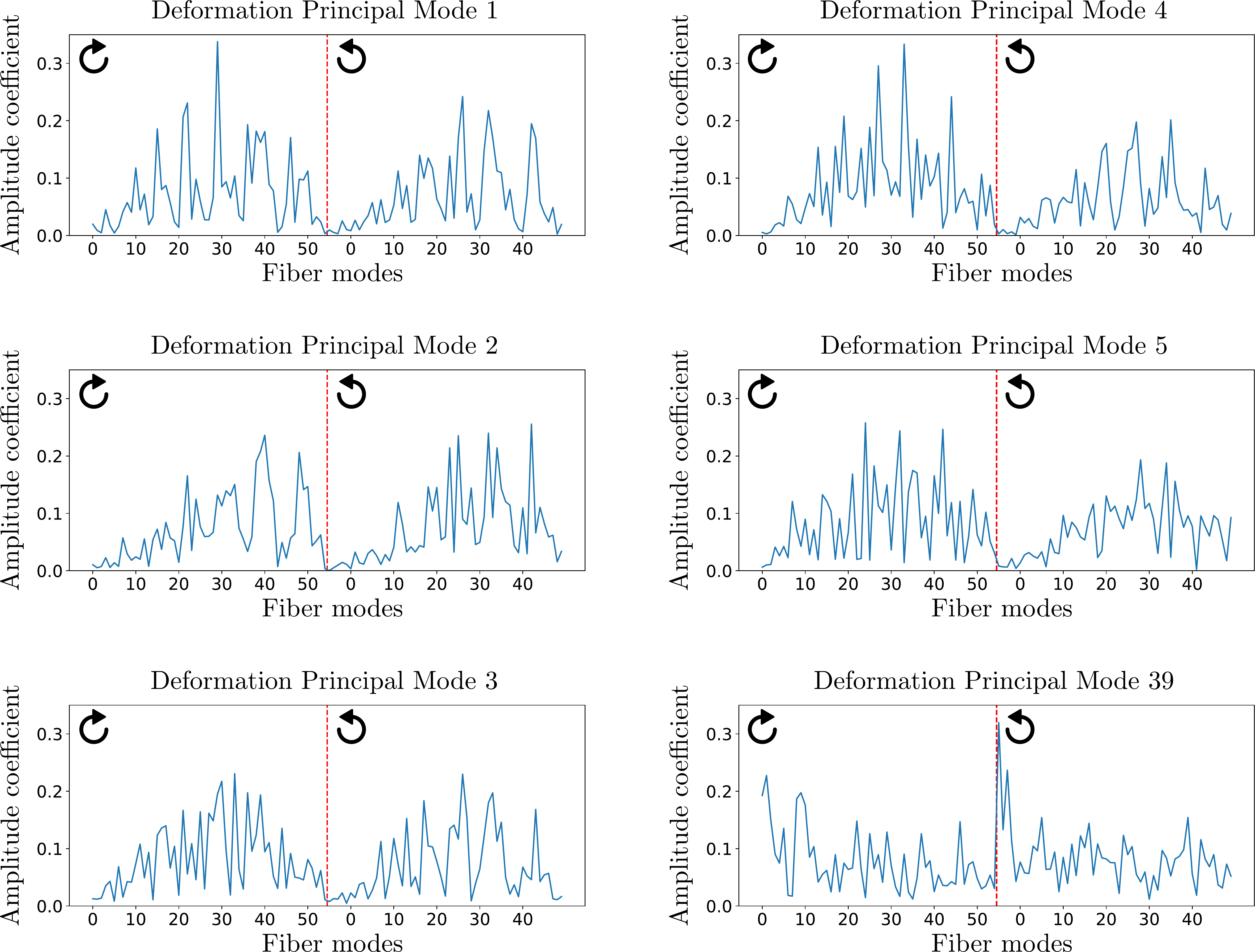}
  \caption{\textbf{Projection of the deformation principal modes onto the fiber modes.}
  Amplitude of the coefficients on the fiber mode basis 
  of the first 5 principal modes and of the one that gives the best correlation between the output profiles at minimal and at maximal deformation (the 39\textsuperscript{th}).
  }
\end{figure}

